\documentclass[prd,a4paper,tightenlines,eqsecnum,nofootinbib,preprint,floatfix,showpacs]{revtex4}
\usepackage{epsfig}
\usepackage{graphicx}
\usepackage{bm}

\def\Tr{\text{Tr}\,}

\def\Eq#1{Eq.~(\ref{#1})}
\def\eff{\text{eff}}

\def\<{\langle}
\def\>{\rangle}
\def\MF{mean-field theory }

\begin{document}

\vspace*{1.0in}

\title{In search of a Hagedorn transition in \\ $SU(N)$ lattice gauge theories at large--$N$
\vspace*{0.25in}}
\author{Barak Bringoltz and Michael Teper\\
\vspace*{0.25in}}

\affiliation{Rudolf Peierls Centre for Theoretical Physics,
University  of  Oxford,\\
1 Keble Road, Oxford, OX1 3NP, UK \\
\vspace*{0.6in}}

\begin{abstract}
We investigate on the lattice the metastable confined phase above $T_c$ in
$SU(N)$ gauge theories, for $N=8,10$, and $12$. In particular
we focus on the decrease with the temperature of the mass of the
lightest state that couples to Polyakov loops. We find that at $T=T_c$
the corresponding effective string tension $\sigma_{\eff}(T)$ is
approximately half its value at $T=0$, and that as we increase $T$ beyond
$T_c$, while remaining in the confined phase, $\sigma_{\eff}(T)$
continues to decrease. We extrapolate $\sigma_{\eff}(T)$ to even
higher temperatures, and interpret the temperature where it vanishes
as the Hagedorn temperature $T_H$. For $SU(12)$ we find that
$T_H/T_c=1.116(9)$, when we use the exponent of the three-dimensional
XY model for the extrapolation, which seems to be slightly preferred over a mean-field exponent by our data.

\end{abstract}

 \maketitle

\section{Introduction}
\label{sec:intro}

In the large--$N$ limit the confined phase of the $SU(N)$ gauge
theory becomes weakly interacting and relatively simpler
\cite{tHooft:1973jz,Witten:1979kh}. Moreover, some of its features
can be described by a low energy effective string theory (for
example see \cite{Polchinski:1992vg}). The link between gauge
theories and string theories was strengthened with the conjectured
AdS/CFT dualities between supersymmetric Yang-Mills theories in the
large--$N$ limit and gravity models (for a review see
\cite{Aharony:1999ti}). An interesting element of systems with
stringy properties is the occurrence of an ultimate temperature,
above which the free string description is unsuitable. In string
models for QCD (or pure gauge theory), this temperature is naturally
identified with a `Hagedorn' deconfinement transition.

The earliest evidence for the existence of a finite temperature
phase transition in hadronic physics was obtained by Hagedorn in
\cite{Hagedorn:1965st}. There, he assumed that hadronic states with
mass $m\to \infty$, referred to as ``fireballs'', are compound systems
of other fireballs. This can be stated mathematically by a
self-consistent relation that the density of states $\rho(E)$ must
obey, and that results in an exponential growth of $\rho$ with the energy $E$.
An immediate consequence is that there exists a temperature $T_H$,
above which the partition function diverges: as $T\to T_H^{-}$, the
Boltzmann suppression of states $\propto \exp(-E/T)$ is overwhelmed
by the density of states, $\rho(E)\propto \exp(+cE)$. At this $T$
an increase of the system's energy is met with an increase in the
number of particles.
In the bootstrap dynamical framework, which predated
the discovery of QCD
and the idea that hadrons are confined bound states of quarks and
gluons, $T_H$ represents an ultimate temperature beyond which
matter cannot be heated. In the more appropriate language of QCD, once
we are close enough to $T_H$ for these fireballs to be densely packed,
the underlying quark and gluon degrees become liberated from
their `hadronic bags' and we expect to see a deconfinement transition \cite{Cabibbo:1975ig}.

A simple, intuitive, and general analogue of the above argument \cite{Polyakov:1978vu,Banks:1979fi}
in the context of any linearly confining theory is as follows.
In such a theory the energy of a string of length $l$
between two distant static sources a distance $r < l$ apart, obeys
\begin{equation}
E(l;r)\simeq \sigma l, \label{Estring}
\end{equation}
where $\sigma$ is the confining string tension and we neglect
subleading terms. For a string (or for a
long enough flux tube) it is easy to see that  once $l\gg r$ the
number of different states of the string grows exponentially with $l$,
$n(l)\propto \exp(cl)$, up to power factors, with $c$ determined
by the dynamics and dimensionality. On the other hand the probability
of such a string with $l\gg r$ is suppressed by a Boltzmann factor,
$\propto \exp(-E(l)/T)$. The total probability of such a string is
given by the free energy that combines the two factors to define an
effective string tension $\sigma_{\rm eff}(T)$:
\begin{equation}
\propto  \exp{\left[ \left( c-\sigma/T\right)\cdot l \right]}
\equiv \exp \left(-\sigma_{{\rm eff}}(T)l/T \right).
\label{eq:stringH}
\end{equation}
It is thus clear that as $T \to T_H \equiv \sigma/c$, arbitrarily
long loops will be thermally excited and the effective string tension
vanishes, $\sigma_{{\rm eff}} \stackrel{T\to T_H}{\longrightarrow}
0$. Because of this vanishing energy, and other features of the free energy \cite{Cabibbo:1975ig}, one would 
naturally expect this
deconfining phase transition to be second order.

We can give a parallel argument for the partition function itself.
Consider an $SU(N)$ gauge theory in which the confining energy
eigenstates are composed of glueballs. A model for glueballs is
to construct them from closed loops of (fundamental) flux. For the
lightest glueballs this loop will be small, and given that the
width of the flux tube is $O(1/\surd\sigma)$, it does not make much
sense to think of a distinct closed loop of flux. However in the
sector of highly excited glueballs, the loop will be very long
and the presence of such states becomes compelling. For a state
composed of a flux loop of length $l$, the energy is $E \sim \sigma l$.
What is the number density, $n_G(E)$, of such states?
Such highly excited states have large quantum numbers and in this
limit a classical counting of states is justified. Thus the number
of these states equals the number of closed loops of length $l$,
i.e. $n_G(E) \propto \exp (cl)$ up to subleading factors, with
the scale $c$ identical to the one in \Eq{eq:stringH}.
Thus the partition function will have a Hagedorn divergence as
$T\to T_H$ where $T_H$ is identical to the string condensation
temperature of the previous paragraph.

Since we expect $c \sim \surd\sigma$, we expect the Hagedorn
transition to occur at $T_H=\sigma/c \sim \surd\sigma$. On the
other hand the lightest glueball (a scalar)  satisfies
$m_G \simeq 3.5 \surd\sigma$ while the next lightest glueball
(a tensor) satisfies $m_G \simeq 4.8 \surd\sigma$. Thus one
has the somewhat counterintuitive picture that as $T\to T_H$
the lightest glueballs are not thermally excited. Rather it
is the highly excited glueballs, whose density
of states grows exponentially with their mass, that drive the
transition.

In the above scenarios, as $T\to T_H$ the vacuum becomes
increasingly densely packed with the thermally excited states.
These will at some point start to interact and the idealised
arguments we use necessarily break down as $T$ approaches $T_H$.
We note that as $N\to\infty$ for $SU(N)$ gauge theories (or QCD),
interactions between colour singlet states vanish. Thus it is
in this limit that the argument for a Hagedorn transition
becomes most compelling.

The vanishing of $\sigma_{\rm eff}$, and the divergence of the associated
correlation length, suggests that this Hagedorn transition is second
order. To what universality class might it belong?
The high temperature phase has a nonzero vacuum expectation value for
the complex valued Polyakov lines that wind around the Euclidean
temporal torus. This spontaneously breaks the global $Z_N$ symmetry of
the theory. Using universality arguments one can then predict the
critical exponents of the transition \cite{Svetitsky:1982gs}. In
particular, in three spatial dimensions it belongs to the universality
classes of the three-dimensional Ising and XY models for $N=2$ and
$N\ge 4$ (for $N=3$, there is no known universality class)
\cite{Svetitsky:1985ye}. Finally for $N=\infty$ there are studies that
predict mean-field behaviour for the correlation length
\cite{Dumitru:2004gd}. This can be understood as a suppression of the
critical region by powers of $1/N$ (see for example the discussion in
\cite{me_strong}, and references therein). The study in \cite{Aharony:2003sx} also gives
mean-field scaling, however only because infrared divergences are not
seen at the small volume discussed there \cite{Aharony_private}.

The above discussion has so far ignored the contribution to the
partition function that comes
from nonconfined energy eigenstates containing a finite density
of gluons. While such states will be irrelevant at low $T$, the fact
that their entropy grows as $N^2$ while the entropy of confined
states is at most weakly dependent on $N$, means that for large enough
$N$ there must be some $T$ where their free energy will decrease
below that of the confined sector of states. At this point there will
be a phase transition to a (perhaps strongly interacting) `gluon
plasma', which one would naturally expect to be first order.
Indeed it turns out to be the case that in four dimensions $SU(N)$ gauge
theories go through a first order deconfining transition for $N\ge 3$
\cite{Lucini:2003zr}. (See also the latent heat calculation at
large-$N$ on a symmetric lattice \cite{Kiskis:2005hf}.)
For SU(2) the transition is second order, but it is not clear if it
is a Hagedorn transition. The fact that it is at small rather
than large $N$ makes the case weaker. As does the fact that the $SU(2)$
value of $T_c/\surd\sigma$ lies on the curve that interpolates
through the $N\geq 3$ values
\cite{Lucini:2003zr}.
On the other hand the value of  $T_c/\surd\sigma$ does coincide
with the string condensation temperature of the simple Nambu-Goto
string theory (see below).

The fact that for  $N\ge 3$ the first order deconfining transition
occurs for $T_c < T_H$, would appear to render the Hagedorn
transition inaccessible. However the deconfining transition
is strongly first order at larger $N$, and so
one can try to use its metastability to
carry out calculations in the confining phase for $T> T_c$.
If $T_H$ is not far away, one can then hope to calculate
$\sigma_{\eff}(T)$ over a range of $T$ where it decreases sufficiently
that an extrapolation to $\sigma_{\eff}(T_H)=0$ can be attempted.
In the range of $N$ accessible to us, the interface tension
between confined and deconfined phases increases with $N$ faster
than the latent heat, and this makes the metastability region
larger \cite{Lucini:2005vg} as $N$ grows. Thus such a strategy
has some chance of success, and it is what we shall attempt in this paper.

Our strategy is therefore to begin deep enough in the confined phase
and then to increase the temperature to temperatures $T>T_c$,
calculating the decrease in $\sigma_{\rm eff}(T)$, and extrapolating
to $\sigma_{\rm eff}=0$. We interpret the result of the extrapolation
as the Hagedorn temperature, $T_H$. Nevertheless, since we work with
finite values of $N$ and volume $V$, tunneling probably occurs
somewhere below $T_H$. These tunneling effects and the fact that
as $\sigma_{\rm eff}$ decreases finite volume effects become
important, can make an apriori fit for the functional form of
$\sigma_{\rm eff}(T)$ unreliable. As a result we first perform fits where we fix the
functional behaviour to be
\begin{equation}
m\equiv \sigma_{\eff}/T \stackrel{_{T\to
  T_H}}{_\sim} \left( T_H/T_c-T/T_c \right)^\nu.
\label{eq:mass_function}
\end{equation}
where $\nu=0.6715(3)$ corresponding to 3D XY \cite{Pelissetto:2000ek}
or $\nu=0.5$ corresponding to mean field. The reason for these two
choices is motivated by two conceivable ways in which the
low energy effective loop potential can behave (see below). In addition we also perform fits
where the exponent $\nu$ is a free parameter, constrained to the range
$[0,1]$, and find it to be especially useful for $SU(12)$. The
coefficient $A$ is fitted as well. An additional outcome of this work
is to confirm that  at $T=T_c$ the mass of the timelike flux loop
that couples to Polyakov loops is far
from zero at large-$N$, which confirms that the transition is strongly
first-order. To make this point clear we will present figures of the
effective string tension in units of the zero temperature string
tension $\sigma$. As a function of $T$, it should have the following
behavior.
\begin{equation}
\sigma_{\eff}/\sigma\stackrel{T\to T_H}{_=}A\cdot T/T_c \cdot \left( T_H/T_c-T/T_c
\right)^\nu,
\label{eq:fit}
\end{equation}

If we imagine the effective potential for an order parameter
such as the Polyakov loop, $V_{\eff}(l_p)$, then at $T=T_c$ this
will possess degenerate  minima corresponding to the confined
and deconfined phases. These will be separated by a barrier whose
height is expected to be $O(N^2)$ at large $N$. As we increase $T$
the confined minimum rises relative to the deconfined one(s).
As $T\to T_H$ we expect the second derivative at the
confining minimum to go to zero,
corresponding to $\sigma_{\eff} \to 0$. The simplest possibility
is that at this $T$ the confining minimum completely disappears,
i.e. that this corresponds to a spinodal point of the potential.
In a string model of glueballs where the glueball is composed
of two `constituent' gluons joined by an adjoint flux tube,
string condensation would correspond to the explicit release
of a gluon plasma simultaneously throughout space and the
identification of the Hagedorn transition with a spinodal
point would be compelling. This is less clear in the closed
flux loop model. (Whether the increasing stability of the
adjoint string as $N\to\infty$ allows us to use the adjoint
string model for the highly excited states relevant near $T_H$
also needs consideration.) If this Hagedorn transition-spinodal point
identification is indeed correct, then in the vicinity of $l_p\simeq
0$, the effective loop potential looks like that of a
Gaussian model, and $\nu=0.5$ is expected. A model for this behaviour is in \cite{Aharony:2003sx}, where 
because of the
infinitesimal volume, any infra-red divergences are excluded, and one
has mean-field scaling close to $T_H$, which also implies $\nu=0.5$.

In principle it seems quite possible that $T_H$
does not coincide with the spinodal transition temperature,
$T_s$. The temperature $T_H$ is a natural concept if one has a good
description of the confined phase as an effective string theory. The
latter will have $O(1/N)$ interactions, and thus to leading order,
lead to a Hagedorn behaviour at a certain temperature $T_H$. On the contrary, $T_s$ most
probably encodes information on the gluonic deconfined phase, which might not be
contained in the string theory. Without any other information in
the spirit of the calculations \cite{Aharony:2003sx}, that identifies $T_H=T_s$,
these two temperatures may be different.

 If $T_H < T_s$ then our method for identifying
$T_H$ remains valid. In that case one may write down a Landau-Ginzburg
theory that has a second order Hagedorn transition at $T_H$,
from the confined vacuum, $C$, to a deconfined one, $D_1$. This
happens at $T_H>T_c$, where both $C,D_1$ are metastable, and another
deconfined vacuum $D_2$ is stable. At large-$N$, the metastability may
be strong enough such that this embedded second order phase transition
happens without being sensitive to tunnelings into the real vacuum. In this
case, the fixed point that controls the critical behavior of the
transition is that of the 3D XY model. Nonetheless, to see this nontrivial critical behavior one must be very 
close to
$T_H$. This happens because interactions between the
Polyakov loops are $O(1/N^2)$ suppressed, and taking the $N=\infty$
limit before $T\to T_H$ results in a Gaussian model and to $\nu=0.5$. To see the scaling behavior
of the 3DXY model, one needs to take $T\to T_H$ together with $N\to
\infty$. In other words, the critical region of this second
order transition has a width that shrinks with increasing $N$
\cite{me_strong}. To see this we can examine the renormalization of
the $\lambda |l_p|^4$ interaction in the Landau-Ginzburg-Wilson effective action of
the transition. As usual, $\lambda$ gets renormalized by infra-red (IR)
modes, and in three spatial dimensions it gets a contribution of
$O(\lambda^2/m)$. Since $\lambda\sim 1/N^2$, this becomes significant
only if $m \sim 1/N^2$. Using $m^2\sim (T_H-T)$, this means that the
IR modes drive the system to the 3DXY universality class only if
$(T_H-T)\stackrel{<}{_\sim}1/N^4$. Outside of this regime the
correlation length has the MF scaling $\nu=0.5$.

Finally, in the case of
$T_H > T_s$ then the spinodal
 transition may (but need not) interfere with our
determination of $T_H$. This is a significant ambiguity that we
cannot resolve in the present calculations but the reader should
be aware of its existence.

We finish by listing some of the reasons that motivate our study.
First it gives nonperturbative information about the  free energy
as a function of the Polyakov loop. While the value of $T_c$ tells us
when the ordered and disordered minima have the same free energy,
the function $\sigma_{\rm eff}(T)$ indicates how the curvature of the
free energy at the confining vacuum changes with $T$ and
its vanishing may indicate the point at which the
confining vacuum becomes completely unstable. This information can
serve to constrain the form of the potential in effective models
for the Polyakov loops (like those in \cite{Dumitru:2004gd}, for example).
Second, this study is a first attempt to investigate the validity of
nonzero temperature \MF techniques of large--$N$ lattice gauge
theories in the continuum. A related study
\cite{Chandrasekharan:2004uw} in the strong-coupling limit saw that
\MF predictions at finite temperature are simply incorrect, which
is consistent with the fact that the large--$N$ limit of these
strongly-coupled fermionic systems is mapped to classical systems at finite
temperature, whose universality class is {\em not} mean-field
\cite{me_strong}. It is of prime interest to know if the same
happens in the continuum limit of the pure gauge theory, which we are much closer to.
The Hagedorn temperature $T_H$ can also give an estimate of the
central charge of a possible underlying string theory
\cite{Meyer:2004hv}, and finally it is interesting to know what
is the limit of $T_H/T_c$ when $N\to \infty$. This limit should
be larger than $1$, given that the deconfining phase transition
has been shown to be first order at $3\leq N\leq 8$ \cite{Lucini:2003zr}.
It is however possible that  $T_H/T_c$ is close to $1$ and perhaps
decreasing with $N$ which opens up interesting new possibilities in
the $N=\infty$ limit.

\section{Lattice calculation}
\label{sec:lattice}

We work on a lattice with $L^3_s\times L_t$ sites, where $L_{t}$ is
the lattice extent in the Euclidean time direction.
The partition function is give by
\begin{equation}
Z=\int DU \exp{\left( -S_{\rm W}\right)},
\end{equation}
where $S_{\rm W}$ is the Wilson action given by
\begin{equation}
S_{\rm W}=\beta \sum_P \left[ 1- \frac1N {\rm Re}\Tr{U_P} \right].
\end{equation}
Here $\beta=2N^2/\lambda$, and $\lambda=g^2N$ is the 't~Hooft
coupling, kept fixed in the large--$N$ limit. $P$ is a lattice
plaquette index, and $U_P$ is the plaquette variable given by
multiplying link variables along the circumference of a fundamental
plaquette. Simulations are done using the Kennedy-Pendleton heat bath
algorithm for the link updates, followed by five over-relaxations. We
focus on the measurement of the correlations of Polyakov lines, which
are taken every five sweeps.

The Monte-carlo simulations were done
using two different methods, which for convenience we refer to as method
``A'' and method ``B''. In both, the calculation was initialized with a
field configuration at the lowest value of $\beta$ which was preceded
by at least $3000$ thermalization sweeps from a cold start. Then,
in method B,
to reduce thermalization effects, each value of $\beta$ was simulated
beginning from the previous one. In contrast, in method A, more than
one simulation began from the same $\beta$.  This serves to eliminate
the dependence between different simulations\footnote{For a {\em finite} amount of statistics these two
methods can lead to different results. In particular, method 'A' should be more noisy.}. We choose to use 
method $A$ and method
$B$ for $N=8$ and $N=12$ respectively. In the case of $N=10$ we used
both methods, and could compare between them. In
Tables~\ref{table1}-\ref{table2}, where we present results obtained
with method A, we give the initial configurations for each value of
$\beta$, along with the statistics of the study. The number of
thermalization sweeps for $N=8,10$ for each value of $\beta$ was always
at least $3000$. As mentioned, the $SU(12)$ calculations were done with
method B, and less thermalization sweeps (at least $400$) were needed
for
each value of $\beta$. For each data set, we calculate the correlation
functions of the thermal lines with improved operators
\cite{Lucini:2005vg}, and use a variational technique to extract the
lightest masses \cite{var0,var1,var2,var3,Luscher:1990ck}, the results
for which are given in lattice units, $am$, in
Tables~\ref{table1}-\ref{table4}, where the errors are
evaluated by a jack-knife procedure.

To check that the results of method A are properly thermalized we also calculate the masses by excluding an 
additional $3000$ sweeps (in addition to the standard thermalization sweeps). This results in a new set of 
masses, which we refer to  as $am_{\rm hot}$, and which we present in Tables~\ref{table1}--\ref{table2}. We 
find that in general the thermalization effects are not significant, and in almost all cases $am_{\rm hot}$ 
agrees with $am$ to within one sigma.

The physical scale $a\surd\sigma$ listed in the Tables was fixed
using the interpolation for the string
tension given in \cite{Lucini:2005vg} in the case of $SU(8)$. For
$N=10,12$ we extrapolate the parameters of the scaling function of
\cite{Lucini:2005vg}, $c_{0,1,2}$, and $\beta_0/N^2$, in $1/N^2$ from their values at $N=6,8$. In addition we
measured the string tension for $N=10$ at $\beta=68.80$
on an $8^4$ lattice, and for $N=12$ at $\beta=99.2,100.0$, on an $8^4$, and a $10^4$ lattice respectively. 
The results, together with the string tensions calculated with the scaling function, are given in 
Table~\ref{table5}.
We find that assuming an error only in the
measured string tension, then the measured and calculated string
tensions deviate at most by $1.6\sigma$. Evaluating the extrapolation error will make the agreement better.

Let us note here that the points used for the fits exclude values of
$\beta$ in which tunneling to the deconfined phase was
observed. Tunnellings were identified by observing double peaks in the
histogram of the expectation values of the Polyakov lines, and by
identifying clear tunneling configurations (again using the Polyakov
lines as a criterion). For other values of $\beta$, we observe an
increase in the fluctuations of the lines with $T$, but did not see
any tunneling configurations. Correspondingly, the histograms widen
with $T$, and for some values of $\beta$ start to develop asymmetric
tails. Nevertheless in all these cases the overall average of the
bare Polyakov line was always lower than $3\times 10^{-3}$.
The $\beta$ values of the excluded deconfining simulations are
$\beta=44.01, 44.015$ for $SU(8)$, and $\beta=69.10, 69.14, 69.16$,
and $69.17$ for $SU(10)$ (for method A), as noted in
Tables~\ref{table1}--\ref{table2}. For the case of $SU(12)$ the last
point of $\beta=99.95$ was still confining after $10,000$ sweeps. At
the next value of $\beta=100.00$ the $12^3\times 5$ system
already showed signs of instability.

Finally, to check the effect of finite volume corrections, which potentially
increase as the mass decreases towards $T_H$, we perform several
additional calculations of the correlation lengths. In the case of
$N=12$ we performed a single calculation on a $16^3\times 5$ lattice
at the smallest volume with $\beta=99.95$. The mass obtained after
5,000 thermalization sweeps and 10,000 measurement sweeps, is
$am=0.187(14)$, and agrees very well with the result obtained on the
$12^3\times 5$ lattice (both are presented in Table~\ref{table4}). For
$N=10$ in method B, all values of $\beta$ were simulated on both a $12^3\times 5$ lattice, and a $14^3\times 
5$ lattice. The
results are presented in Table~\ref{table3}, and we find at most a
$1.3\sigma$ difference between them. Comparing
with the situation close to the second order phase transition of the
$SU(2)$ group \cite{Lucini:2005vg}, we find that for $N=2$, finite
volume effects are much more important than for $N=12$. This puts the
results of
this work, which were largely done for only one lattice volume, on a more
solid footing. This is also consistent with standard theoretical arguments that predict smaller volume 
corrections for gauge theories with larger values of $N$.

\begin{table}[htb]
\caption{Statistics and results of the Monte-Carlo simulations for $N=8$, obtained with method A. The ``D'' 
denotes tunneling to deconfining configurations, which is not considered for the fits. $am_{\text{ hot}}$ is 
the result of the data set that excludes the first 3000 measurement sweeps.
\label{table1}}
\begin{ruledtabular}
\begin{tabular}{cccccc}
$N=8$ & & & & \\ \hline
 $\beta$  & $am$ &  $am_{\text{ hot}}$  & $a\sqrt{\sigma}$ & Initial & (no. of  sweeps)$/10^3$\\
   & (all sweeps) &   & &   \\ \hline
43.850  &       0.361(17)       &       0.362(17)       &       0.3615          &       Frozen  &       20.0   
\\
43.875  &       0.336(22)       &       0.362(22)       &       0.3580          &       43.850  &       15.0    
\\
43.900  &       0.334(17)       &       0.337(17)       &       0.3547          &       43.875  &       13.0    
\\
43.930  &       0.272(19)       &       0.270(19)       &       0.3507          &       43.875  &       17.0    
\\
43.950  &       0.314(15)       &       0.326(15)       &       0.3481          &       43.850  &       24.0    
\\
43.975  &       0.286(17)       &       0.283(17)       &       0.3448          &       43.950  &       10.0    
\\
43.980  &       0.258(18)       &       0.245(18)       &       0.3442          &       43.975  &       13.0    
\\
43.985  &       0.304(14)       &       0.294(14)       &       0.3436          &       43.975  &       13.0    
\\
43.995  &       0.206(13)       &       0.207(13)       &       0.3423          &       43.975  &       14.0    
\\
44.000  &       0.264(17)       &       0.260(17)       &       0.3417          &       43.850  &       29.0    
\\
44.01;D &                       -               &                       -               &       0.3404          
&       44.000  &       13.0    \\
44.015;D        &                       -               &                       -               &       
0.3398          &       44.000  &       13.0    \\
44.020  &       0.240(17)       &       0.253(17)       &       0.3392          &       44.000  &       15.0    
\\
44.025  &       0.228(17)       &       0.201(17)       &       0.3385          &       44.000  &       16.0    
\\
44.033  &       0.213(14)       &       0.220(14)       &       0.3376          &       44.000  &       20.0    
\\
\end{tabular}\end{ruledtabular}
\end{table}
\begin{table}[htb]
\caption{Results and statistics of the Monte-Carlo simulations for $N=10$ obtained with method A. The ``D'' 
denotes tunneling to deconfining configurations. $am_{\rm hot}$ are the results of the data set that excludes 
the first 3000 measurement sweeps.
\label{table2}}
\begin{ruledtabular}\begin{tabular}{cccccc}
$N=10$ & &  & & & \\ \hline
 $\beta$  & $am$ & $am_{\text{hot}}$  & $a\sqrt{\sigma}$ & Initial  & (no. of   \\
& (all sweeps)&  & & config. & sweeps)/$10^3$  \\ \hline
68.5000 &       0.571(41)       &       0.609(34)       &       0.3941          &       frozen  &       17.0    
\\
68.5520 &       0.524(28)       &       0.514(29)       &       0.3888          &       68.50   &       22.0    
\\
68.6000 &       0.545(11)       &       0.544(11)       &       0.3839          &       68.50   &       23.5    
\\
68.6553 &       0.502(13)       &       0.500(14)       &       0.3785          &       68.60   &       20.5    
\\
68.7000 &       0.481(36)       &       0.441(35)       &       0.3742          &       68.50   &       23.5    
\\
68.7500 &       0.452(12)       &       0.453(14)       &       0.3695          &       68.50   &       23.0    
\\
68.8000 &       0.391(17)       &       0.388(19)       &       0.3649          &       68.50   &       26.0    
\\
68.8500 &       0.389(14)       &       0.378(22)       &       0.3603          &       68.50   &       24.0    
\\
68.9000 &       0.356(15)       &       0.345(16)       &       0.3559          &       68.50   &       23.0    
\\
68.9500 &       0.287(15)       &       0.289(16)       &       0.3516          &       68.50   &       23.5    
\\
69.0000 &       0.295(17)       &       0.286(17)       &       0.3473          &       68.50   &       25.0    
\\
69.0100 &       0.295(13)       &       0.294(14)       &       0.3465          &       69.00   &       23.0    
\\
69.0200 &       0.315(15)       &       0.312(15)       &       0.3457          &       69.00   &       22.0    
\\
69.0412 &       0.277(15)       &       0.292(16)       &       0.3439          &       69.00   &       22.0    
\\
69.0500 &       0.271(14)       &       0.273(16)       &       0.3432          &       69.00   &       21.0    
\\
69.0700 &       0.251(18)       &       0.242(19)       &       0.3416          &       69.00   &       22.0    
\\
69.0816 &       0.262(16)       &       0.258(17)       &       0.3406          &       69.00   &       22.0    
\\
69.1000 &                       -               &                       -               &       0.3391          
&       68.50   &       11.0    \\
69.1100 &       0.253(11)       &       0.248(11)       &       0.3383          &       69.10   &       22.0    
\\
69.1200 &       0.233(15)       &       0.226(16)       &       0.3375          &       69.10   &       22.0    
\\
69.1300 &       0.223(15)       &       0.231(14)       &       0.3367          &       69.10   &       22.0    
\\
69.1400 &                       -               &                       -               &       0.3359          
&       69.10   &       12.0    \\
69.1500 &       0.218(12)       &       0.228(13)       &       0.3351          &       69.10   &       22.0    
\\
69.1600 &                       -               &                       -               &       0.3344          
&       69.15   &       8.0     \\
69.1700 &                       -               &                       -               &       0.3336          
&       69.15   &       8.0     \\
69.1800 &       0.222(14)       &       0.225(16)       &       0.3328          &       69.15   &       18.0    
\\
69.1900 &       0.218(12)       &       0.228(13)       &       0.3320          &       69.15   &       18.0    
\\
\end{tabular}\end{ruledtabular}
\end{table}
\begin{table}[htb]
\caption{Masses in lattice units and statistics of the Monte-Carlo simulations for $N=10$ obtained with 
method B on $14^3\times 5$, and $12^3\times 5$ lattices.
\label{table3}}
\begin{ruledtabular}\begin{tabular}{cccccc}
$N=10$ & & & &  & \\ \hline
 $\beta$  & $12^3\times 5$ & (no. of sweeps)/$10^3$ & $14^3\times 5$ & (no. of sweeps)$/10^3$ & 
$a\sqrt{\sigma}$  \\ \hline
68.50   &       0.599(30)       &       20      &       0.584(35)       &       20      &       0.3941  \\
68.70   &       0.396(17)       &       22      &       0.428(29)       &       22      &       0.3742  \\
68.85   &       0.371(27)       &       20      &       0.371(18)       &       20      &       0.3603  \\
68.95   &       0.316(16)       &       ``      &       0.308(11)       &       ``      &       0.3516  \\
69.07   &       0.246(12)       &       ``      &       0.270(15)       &       ``      &       0.3416  \\
69.13   &       0.244(17)       &       ``      &       0.241(14)       &       18      &       0.3367  \\
69.17   &       0.220(10)       &       ``      &       0.203(15)       &       16      &       0.3336  \\
69.20   &       0.154(11)       &       22      &       0.200(21)       &       8       &       0.3313  \\
69.23   &       0.183(19)       &       15      &                       -               &       -       &       
0.3290  \\
\end{tabular}\end{ruledtabular}
\end{table}
\begin{table}[htb]
\caption{Results and statistics of the Monte-Carlo simulations for $N=12$ obtained with method B. All masses 
are calculated on a $12^3\times 5$ lattice, except for the last row which is for $16^3\times 5$ lattice. The 
number of thermalization sweeps was at least 400 between two successive values of $\beta$, while it was $800$ 
for the first calculation of $\beta=99.00$. \label{table4}}
\begin{ruledtabular}\begin{tabular}{cccc}
$N=12$ & & & \\ \hline
 $\beta$ & $am$   & $a\sqrt{\sigma}$ & (no. of sweeps)$/ 10^3$ \\ \hline
99.00   &       0.573(25)       &       0.3866  &       5       \\
99.20   &       0.472(24)       &       0.3729  &       5       \\
99.40   &       0.396(18)       &       0.3601  &       5       \\
99.60   &       0.333(17)       &       0.3479  &       10      \\
99.80   &       0.259(13)       &       0.3365  &       10      \\
99.90   &       0.203(13)       &       0.3310  &       10      \\
99.95   &       0.190(13)       &       0.3284  &       10      \\
99.95   &       0.187(14)       &       0.3284  &       10      \\
\end{tabular}
\end{ruledtabular}\end{table}
\begin{table}[htb]
\caption{String tensions for $SU(10)$ and $SU(12)$.
\label{table5}}
\begin{ruledtabular}\begin{tabular}{ccc}
$N,\beta$  & Measurement & Results of scaling function (see text) \\ \hline
$N=10,\beta=68.80$   &       0.3667(80)      &       0.3649  \\
$N=12,\beta=99.20$   &       0.3770(26)      &       0.3729  \\
$N=12,\beta=100.00$  &       0.3243(23)      &       0.3257  \\
\end{tabular}\end{ruledtabular}
\end{table}

If we define the effective string tension,  $\sigma_{\eff}(T)$, to be
the coefficient of the leading linear part of the free energy
of two distant fundamental sources, then
$\sigma_{\eff}(T) = mT$ where $m$ is the lightest mass that
couples to timelike Polyakov loops. Therefore the ratio
$\sigma_{\eff}/\sigma$ will be given by
\begin{equation}
\frac{\sigma_{\eff}}{\sigma}=\frac{a m(\beta)}{L_t\cdot
(a\sqrt{\sigma}(\beta))^2}.
\end{equation}
At each simulated ratio of $\beta$ we estimate  $T/T_c$ using
\begin{equation}
\frac{T}{T_c}=\frac{(a\sqrt{\sigma})_c}{a\sqrt{\sigma}(\beta)},
\end{equation}
where $(a\sqrt{\sigma})_c$ is the value of $a\sqrt{\sigma}$ at
$\beta=\beta_c$ for $L_t=5$, i.e. $5a(\beta_c)=T^{-1}_c$.
It is extracted from $T_c/\sqrt{\sigma}$, which is 0.5819(41) for
$N=8$. (This assumes that  $T_c/\sqrt{\sigma}$ varies at most
very weakly with $a(\beta)$, which is in fact what one observes \cite{Lucini:2005vg}.)
The corresponding value for
$N=10,12$ is found by extrapolating in $1/N^2$ according to measured
values of $T_c/\sqrt{\sigma}(L_t=5)$ for $N=4,6,8$
\cite{Mike_email}. This gives 0.5758 for $SU(10)$ and 0.5735 for
$SU(12)$, with an error of about $1\%$.

\begin{figure}[htb]
\includegraphics[width=15cm]{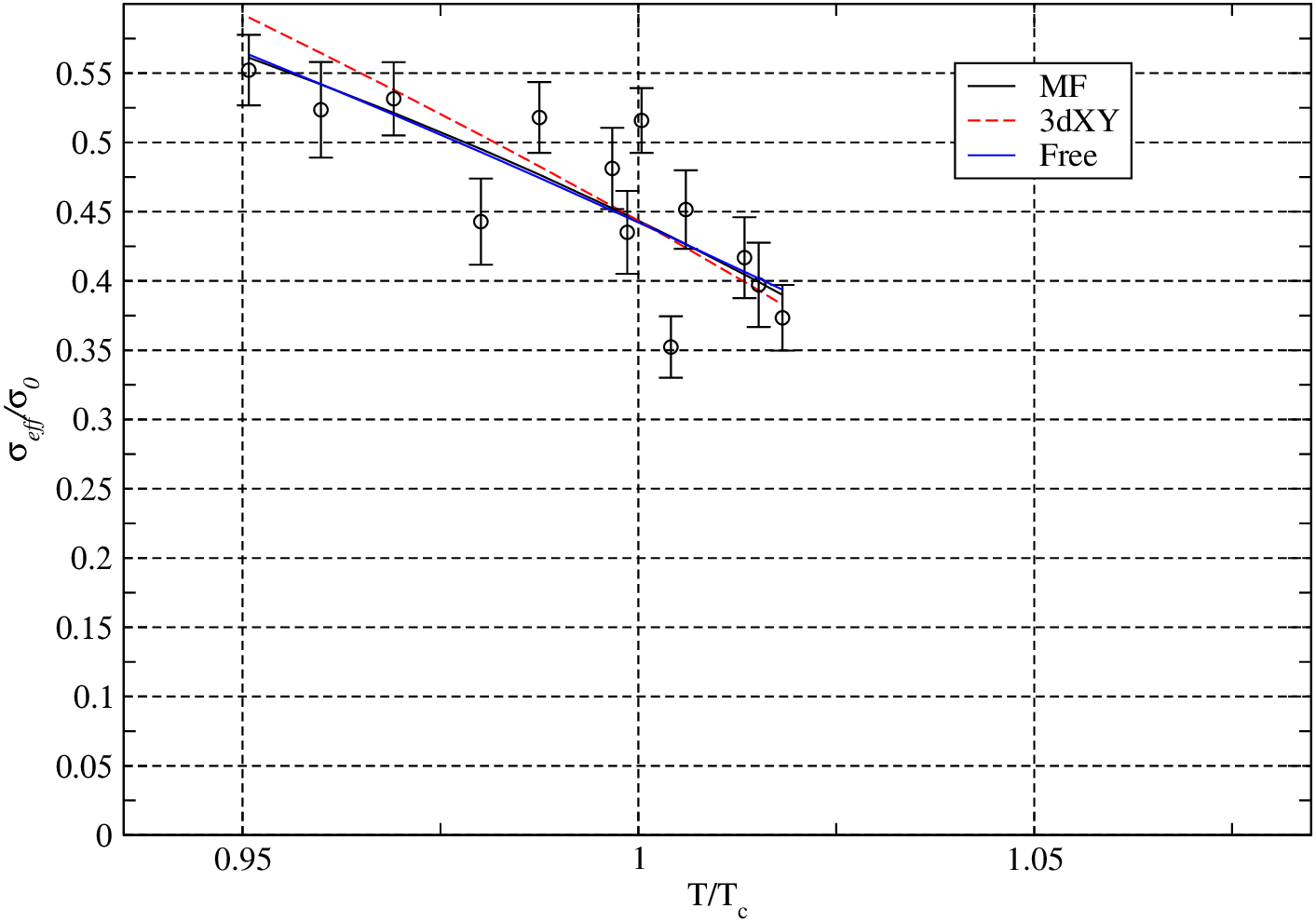}
\vspace{0.5cm}
\caption{Effective string tension for $SU(8)$ obtained in method A in units of the zero temperature string 
tension.}
\label{fig:N8}
\end{figure}

\begin{figure}[htb]
\includegraphics[width=15cm]{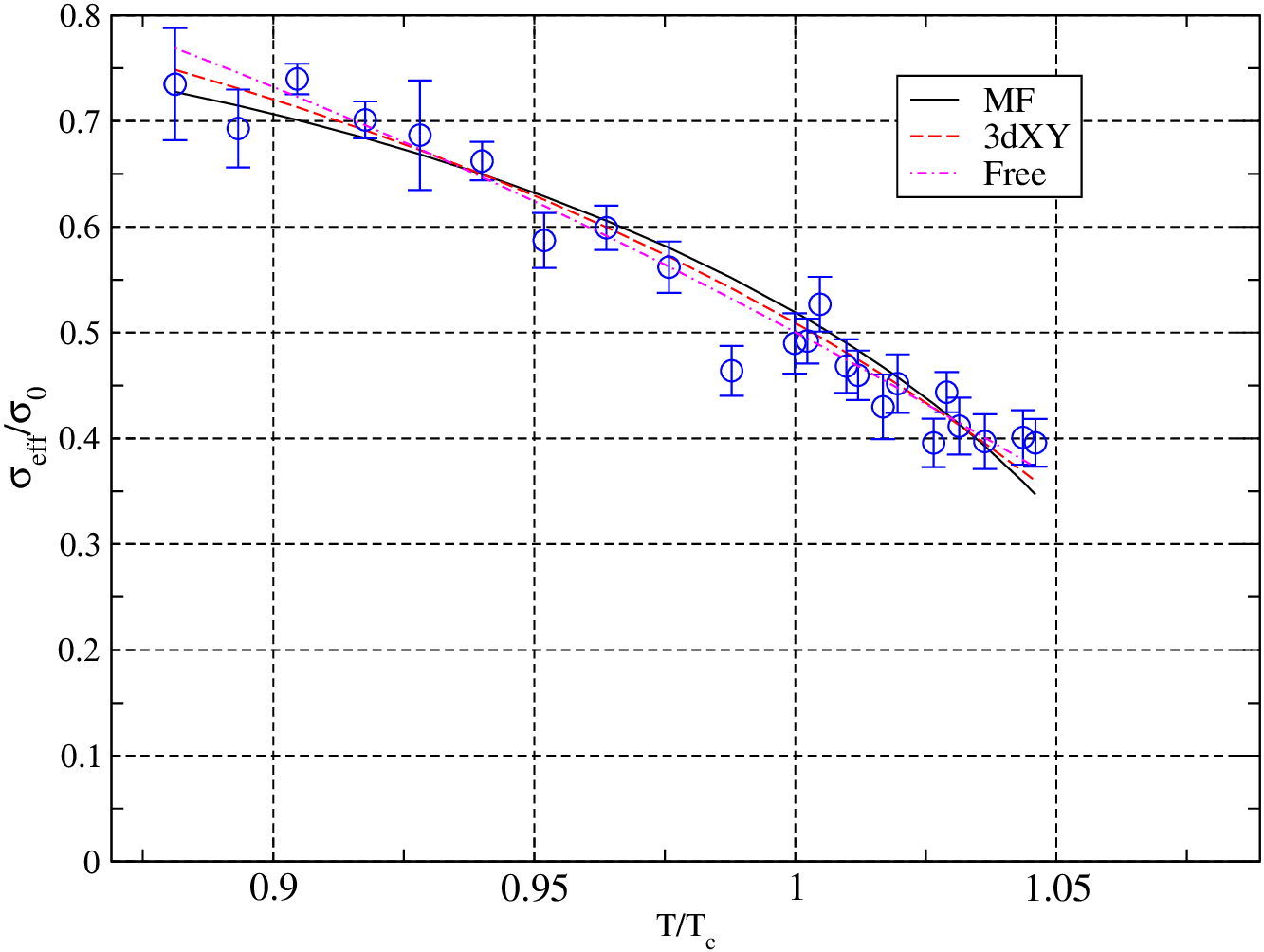}
\vspace{0.5cm}
\caption{Effective string tension for $SU(10)$ obtained in method A in units of the zero temperature string 
tension.}
\label{fig:N10}
\end{figure}

\begin{figure}[htb]
\includegraphics[width=15cm]{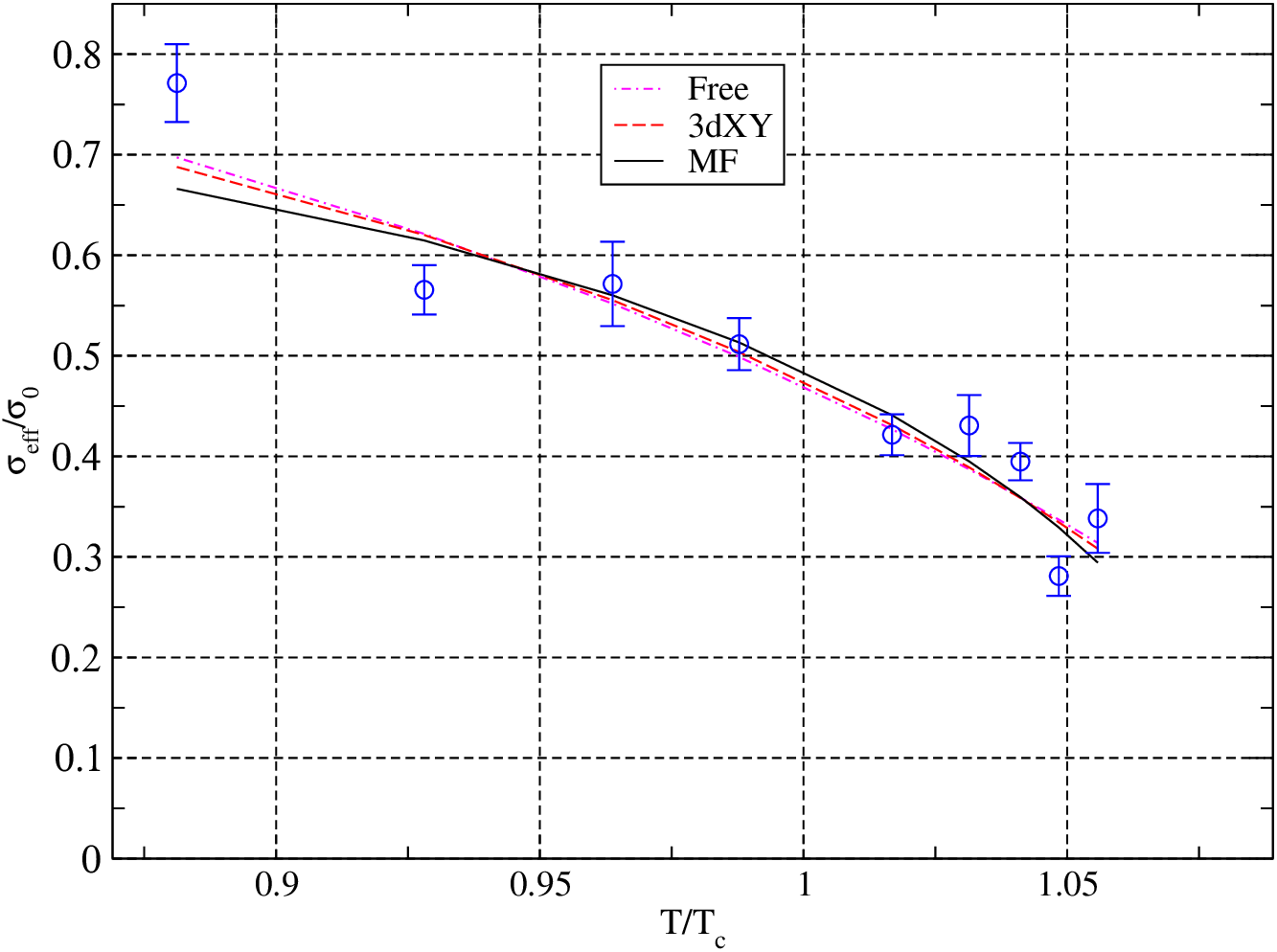}
\vspace{0.5cm}
\caption{Effective string tension for $SU(10)$ obtained in method B on a $12^3\times 5$ lattice, in units of 
the zero temperature string tension.} \label{fig:N10succesiveL12}
\end{figure}

\begin{figure}[htb]
\includegraphics[width=15cm]{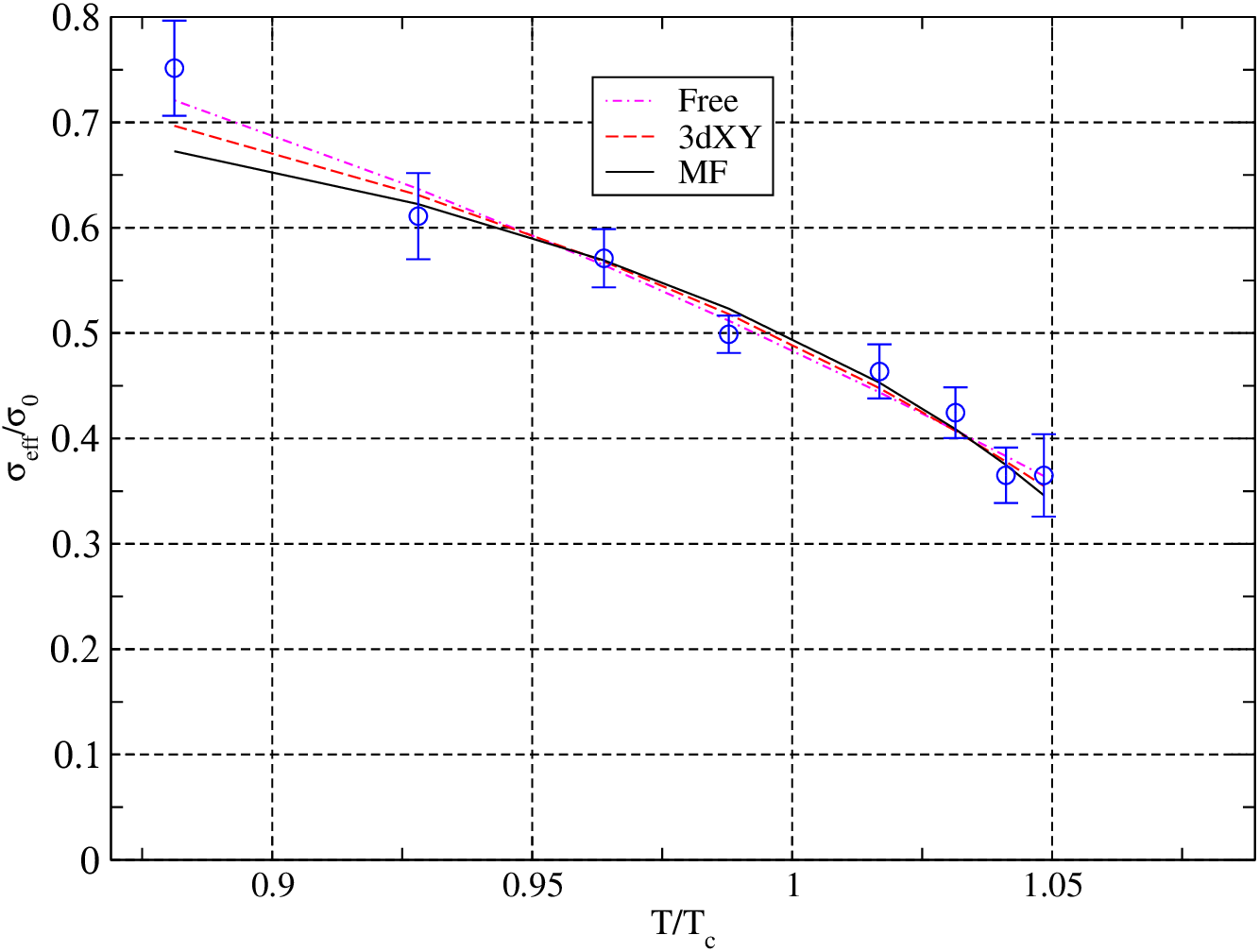}
\vspace{0.5cm}
\caption{Effective string tension for $SU(10)$ obtained in method B on a $14^3\times 5$ lattice, in units of 
the zero temperature string tension.} \label{fig:N10succesiveL14}
\end{figure}

\begin{figure}[htb]
\includegraphics[width=15cm]{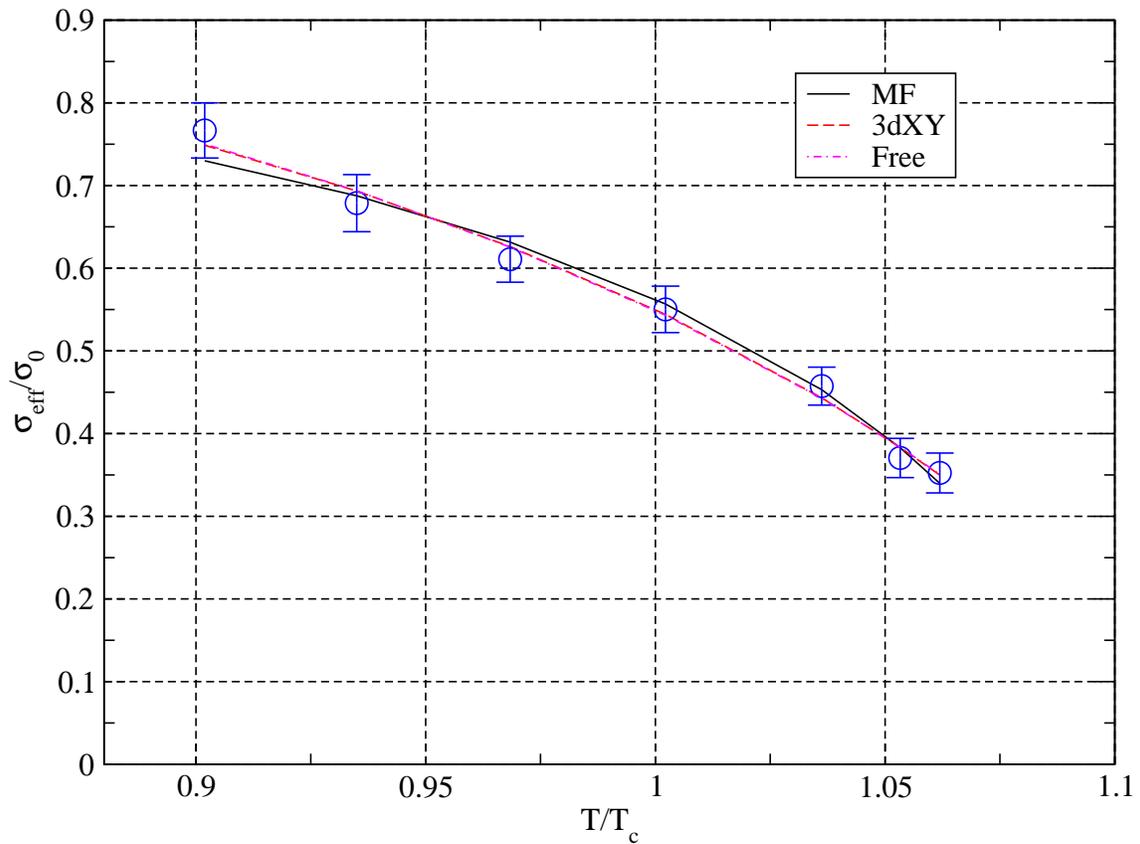}
\vspace{0.5cm}
\caption{Effective string tension for $SU(12)$ obtained in method B in units of the zero temperature string 
tension.} \label{fig:N12}
\end{figure}

In Figs.~\ref{fig:N8}--\ref{fig:N12} we give the
effective string tensions as a function of temperature for the studied
gauge groups and various data sets, as explained above. In view of
\Eq{eq:fit} we choose to present
 $\sigma_{\eff}/\sigma$.
As the relative error of $am$ is roughly 10 times the one on
$T_c/\sqrt{\sigma}$,
we neglect the latter in our error estimate. The fit to the data was done according \Eq{eq:fit}, either by 
fixing $\nu=0.6715,0.5$ for the 3D XY and mean-field universality classes, or in some cases by making $\nu$ a 
free parameter. The fitting results are presented in Table~\ref{table6}.

\begin{table}[htb]
\caption{Results of the fits to \Eq{eq:fit}. The values of $T_H/T_c$, and $A$ are presented for all fits. In 
the cases where a fit with $\nu$ as a free parameter was made, then the resulting $\nu$ is presented in the 
second column. We present the coverage probability contours in the $A-T_H/T_c$ plane, when appropriate, in 
Figs.~\ref{fig:confidence10A}--\ref{fig:confidence12free}.
\label{table6}}
\begin{ruledtabular}
\begin{tabular}{c|ccccccc}
$N$ & Universality class & $\frac{T_H}{T_c}$ & $\frac{T_H}{\sqrt{\sigma}}$ & $A$ & central charge & 
$\frac{\chi^2}{{\text{dof}}}$ & dof \\ \hline
8 & 3D XY   & 1.093 & 0.636 & 2.190 & 1.178 & 2.84 & 11\\
$12^3\times 5$ & MF   & 1.066  & 0.620 & 1.728 & 1.237 &2.79 & 11\\
(method A) & Free, $\nu=1$   & 1.145  & 0.666 & 3.055 & 1.073 &3.19 & 10\\ \hline
10 & 3D XY   & 1.104 & 0.636 & 2.328  & 1.236 & 1.44 & 21\\
$12^3\times 5$ & MF   & 1.078 & 0.621 & 1.862  & 1.178 & 2.07 & 21\\
(method A) & Free, $\nu=1$   &  1.160 & 0.668 & 3.132   & 1.067 & 1.16  & 20\\ \hline
10 & 3D XY   & 1.108 & 0.624 & 2.112  & 1.169 & 3.29 & 7\\
$12^3\times 5$ & MF   & 1.083 & 0.638 & 1.681  & 1.223 & 3.56 & 7\\
(method B) & Free, $\nu=0.79$   &  1.127 & 0.649 & 2.394   & 1.131 & 3.94  & 6\\ \hline
10 & 3D XY   & 1.114 & 0.642 & 2.102  & 1.156 & 0.68 & 6\\
$14^3\times 5$ & MF   & 1.087 & 0.626 & 1.683 & 1.215 & 1.00 & 6\\
(method B) & Free, $\nu=1$   &  1.172 & 0.675 & 2.815   & 1.045 & 0.60  & 5\\ \hline
12 & 3D XY   &  1.116 & 0.640 & 2.337  & 1.162& 0.30 & 5\\
$12^3\times 5$ & MF   &  1.092 & 0.626 & 1.858  & 1.215& 0.50 & 5\\
(method B) & Free, $\nu=0.69$   & 1.119 & 0.642 & 2.387 & 1.156 & 0.37 & 4\\ \hline

\end{tabular}\end{ruledtabular}\end{table}

\begin{figure}[htb]
\resizebox{7cm}{!}{\includegraphics{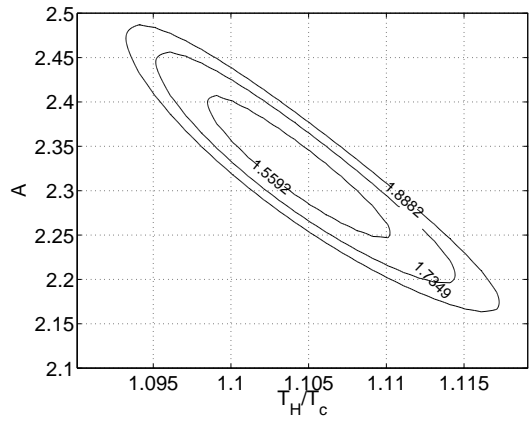}}
\vspace{0.5cm}
\caption{Confidence levels for the fits of $SU(10)$ on a $12^3\times 5$ obtained with method A for the 3DXY 
exponent.}
\label{fig:confidence10A}
\end{figure}

\begin{figure}[htb]
\resizebox{7cm}{!}{\includegraphics{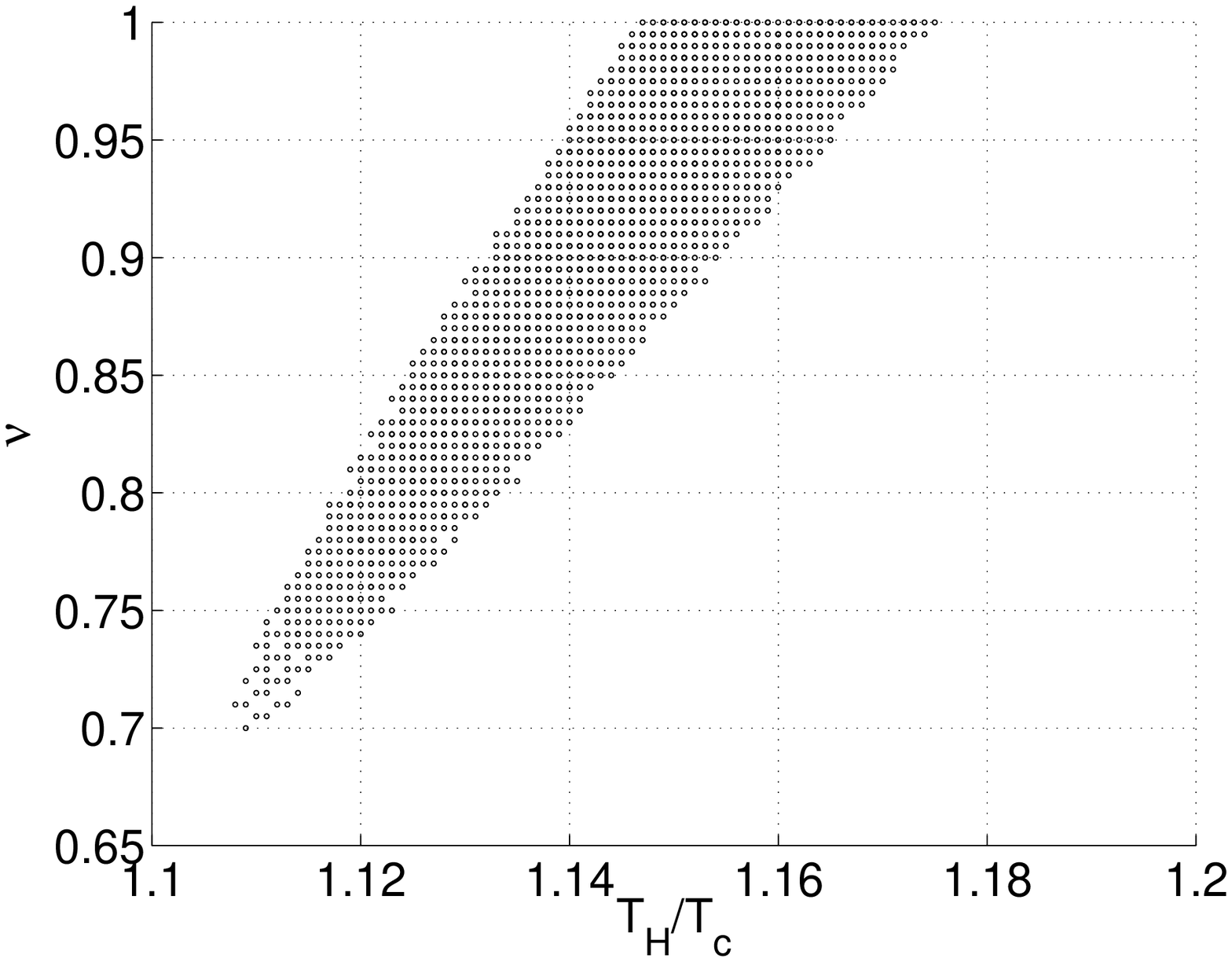}} \qquad
\resizebox{7cm}{!}{\includegraphics{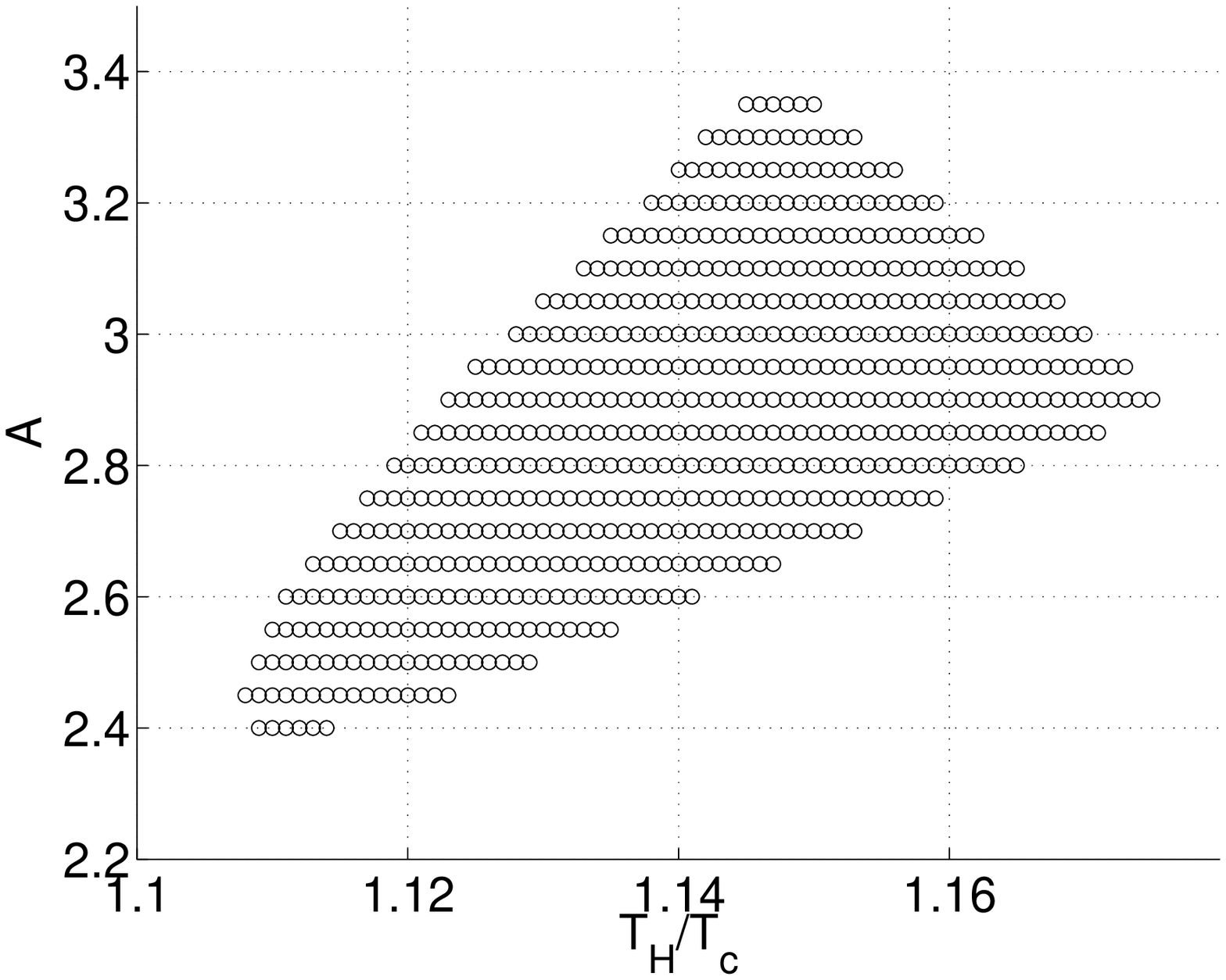}}
\vspace{0.5cm}
\caption{Two dimensional projections of a isosurface in the three-dimensional space of $(\nu,T_H/T_c,A)$ 
which represents a confidence level of $68.27\%$ for the fits of $SU(10)$ obtained with method A on a 
$12^3\times 5$ lattice, with a free exponent.} \label{fig:confidence10AL12free}
\end{figure}

\begin{figure}[htb]
\resizebox{7cm}{!}{\includegraphics{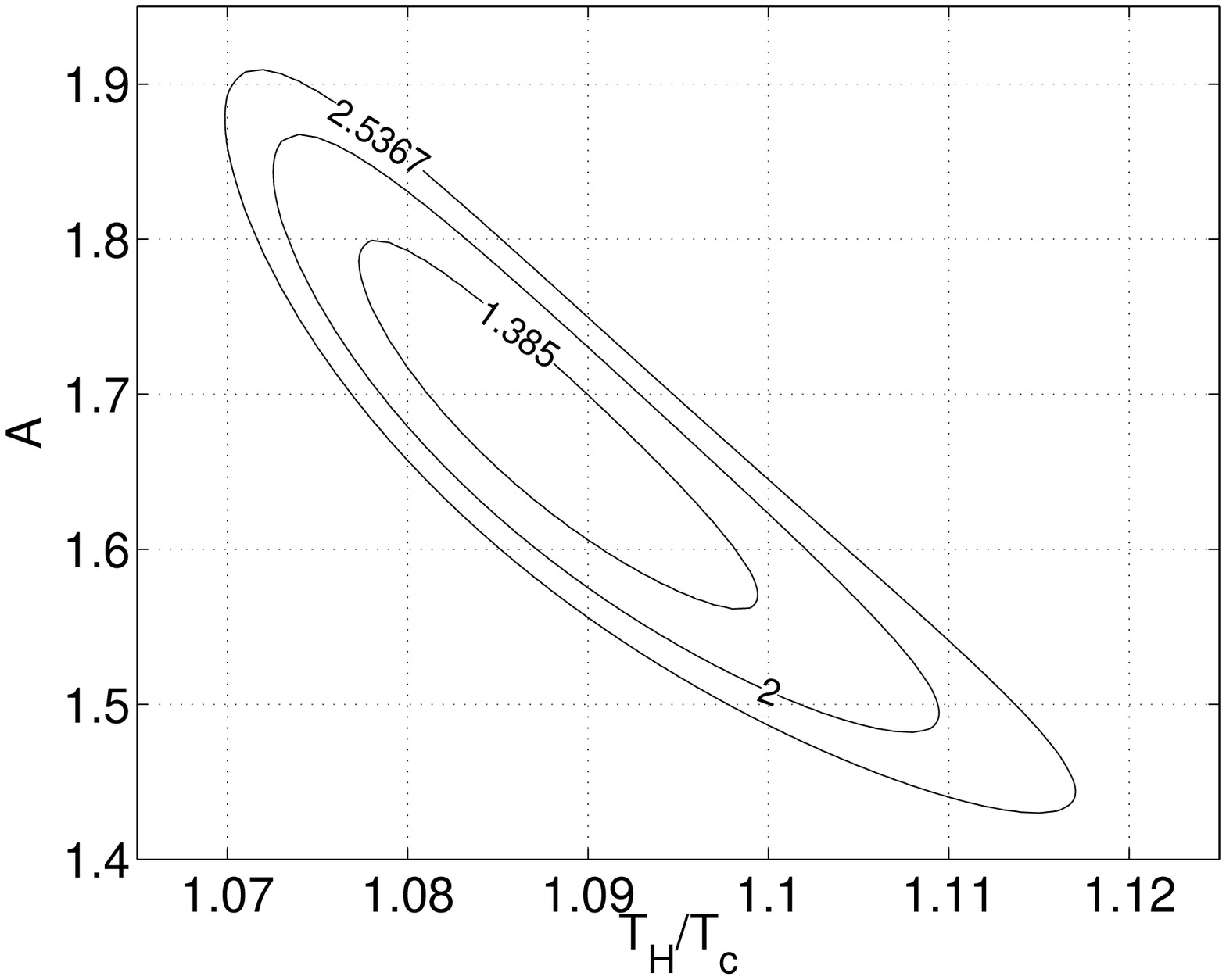}} \qquad
\resizebox{7cm}{!}{\includegraphics{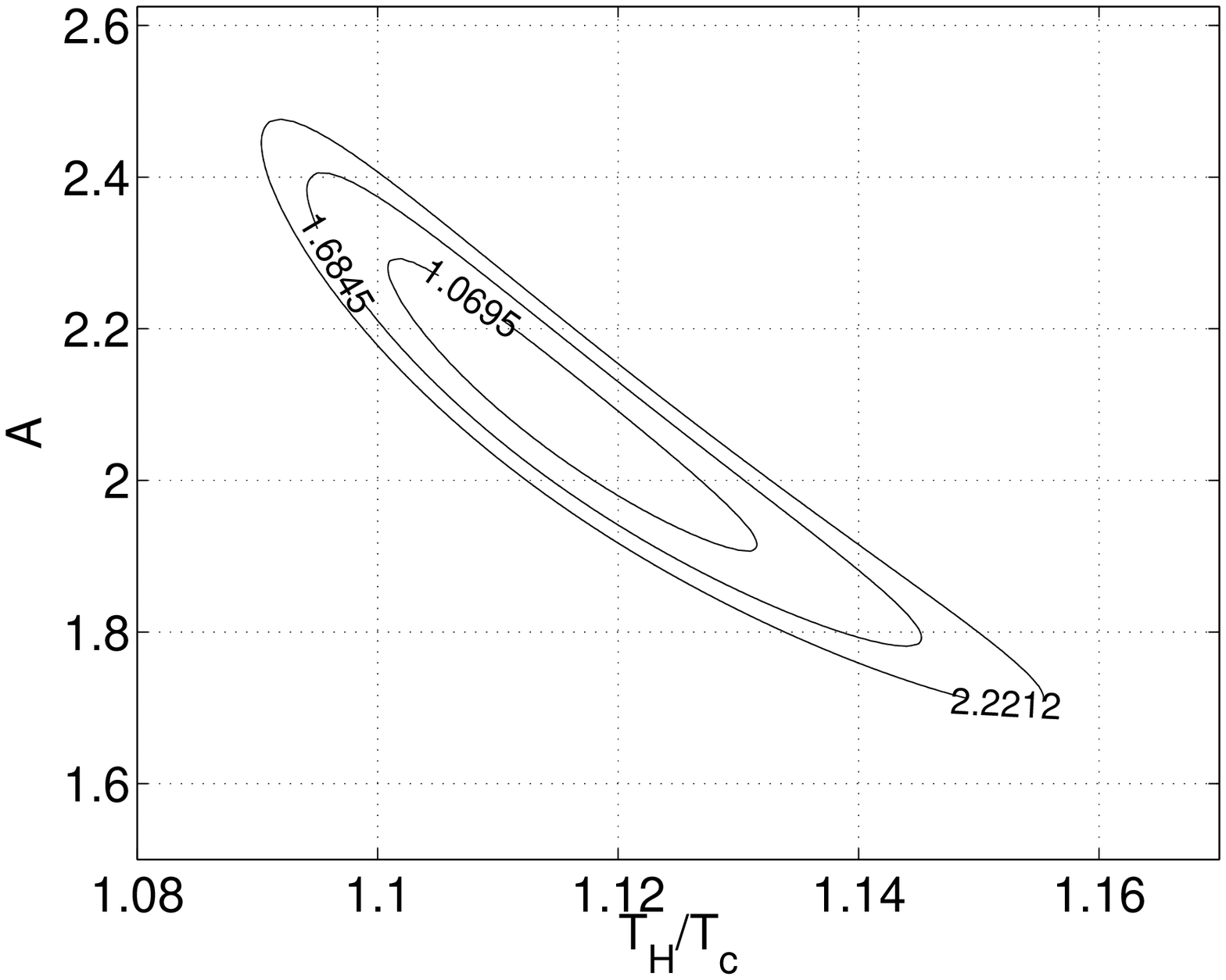}}
\vspace{0.5cm}
\caption{Confidence levels for the fits of $SU(10)$ on a $14^3\times 5$ obtained with method B. The plot on 
the right(left) is for the 3DXY(MF) exponent.} \label{fig:confidence10BL14}
\end{figure}

\begin{figure}[htb]
\resizebox{7cm}{!}{\includegraphics{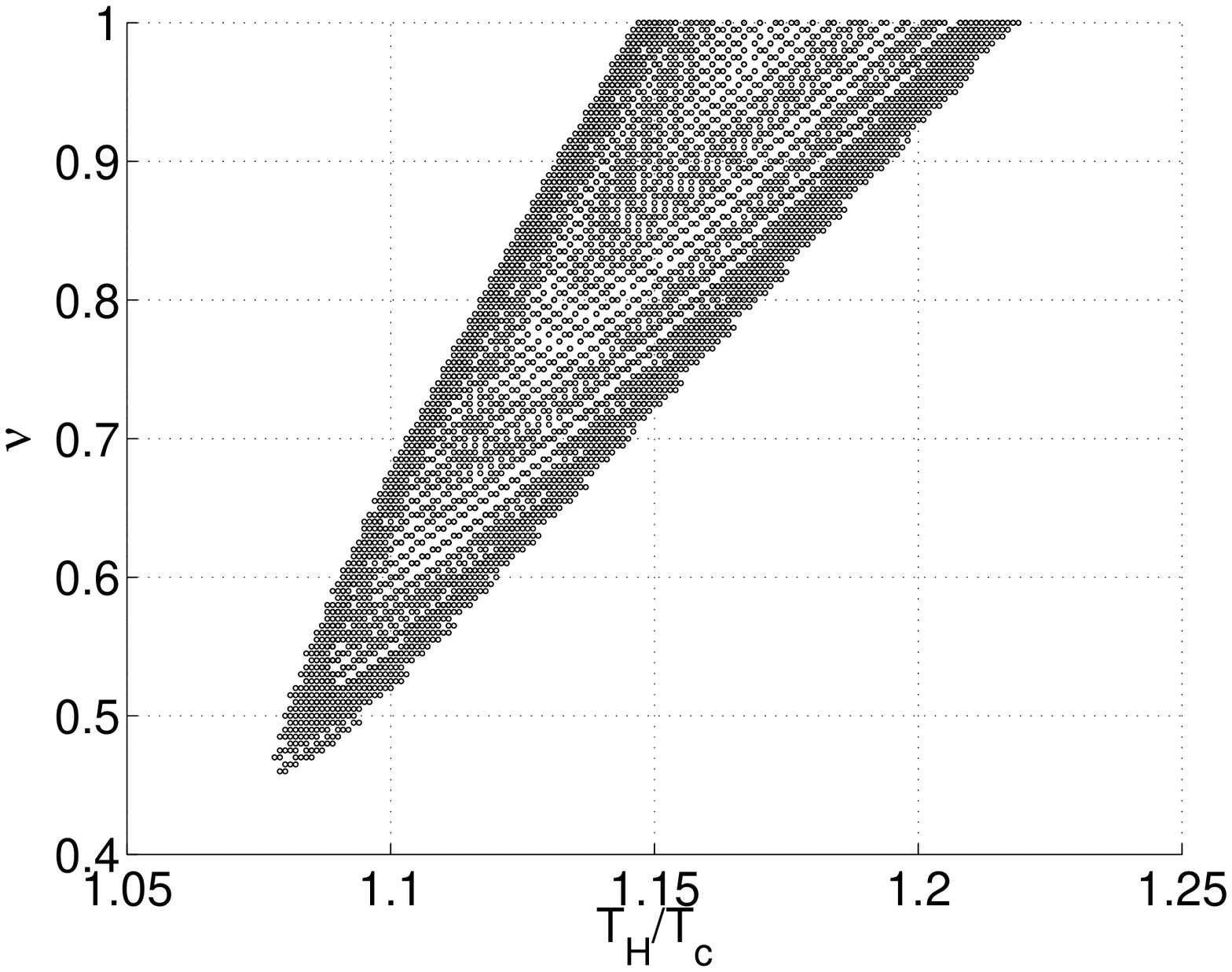}} \qquad
\resizebox{7cm}{!}{\includegraphics{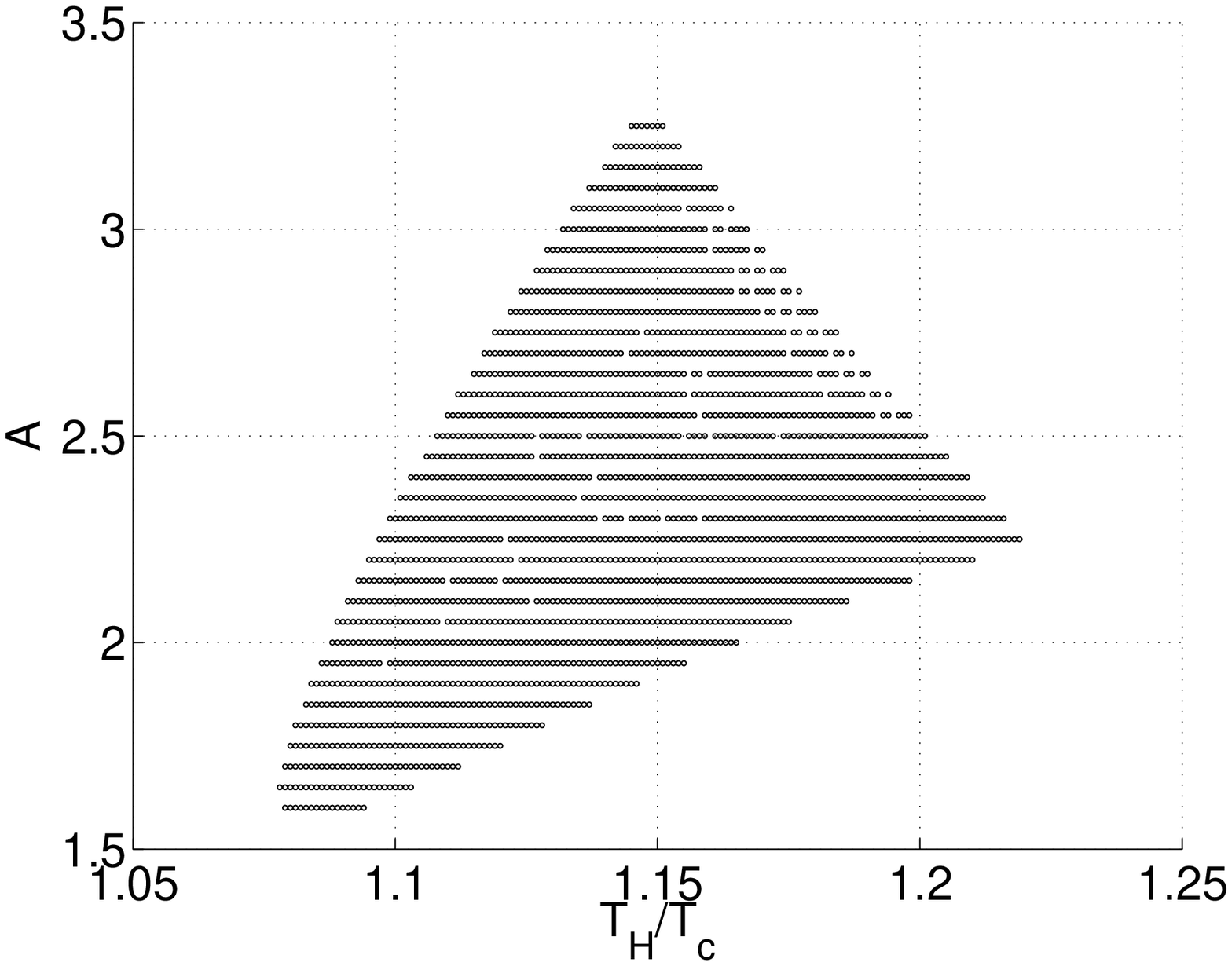}}
\vspace{0.5cm}
\caption{Two dimensional projections of a isosurface in the three-dimensional space of $(\nu,T_H/T_c,A)$ 
which represents a confidence level of $68.27\%$ for the fits of $SU(10)$ obtained with method B on a 
$14^3\times 5$ lattice, with a free exponent.} \label{fig:confidence10BL14free}
\end{figure}

\begin{figure}[htb]
\resizebox{7cm}{!}{\includegraphics{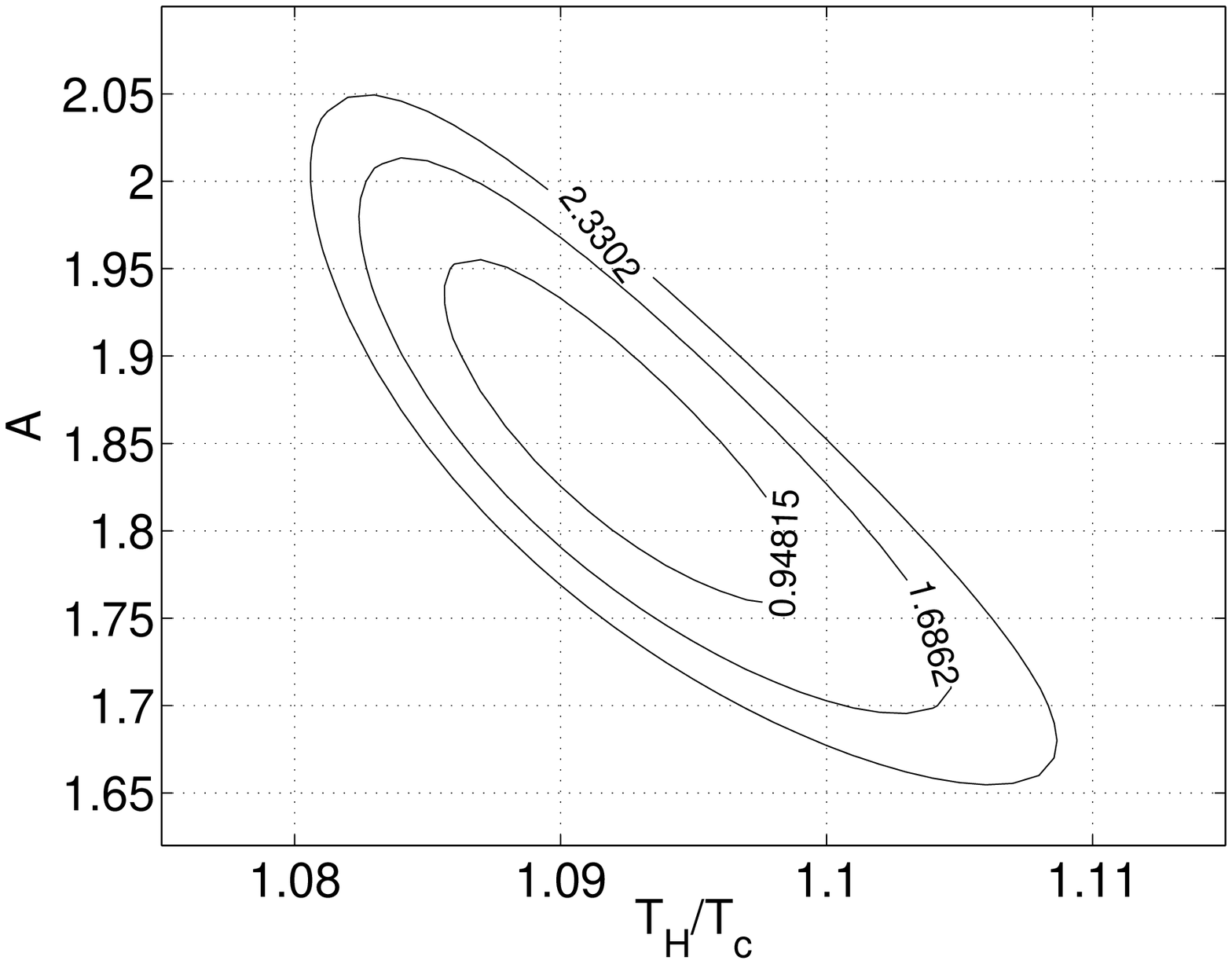}} \qquad
\resizebox{7cm}{!}{\includegraphics{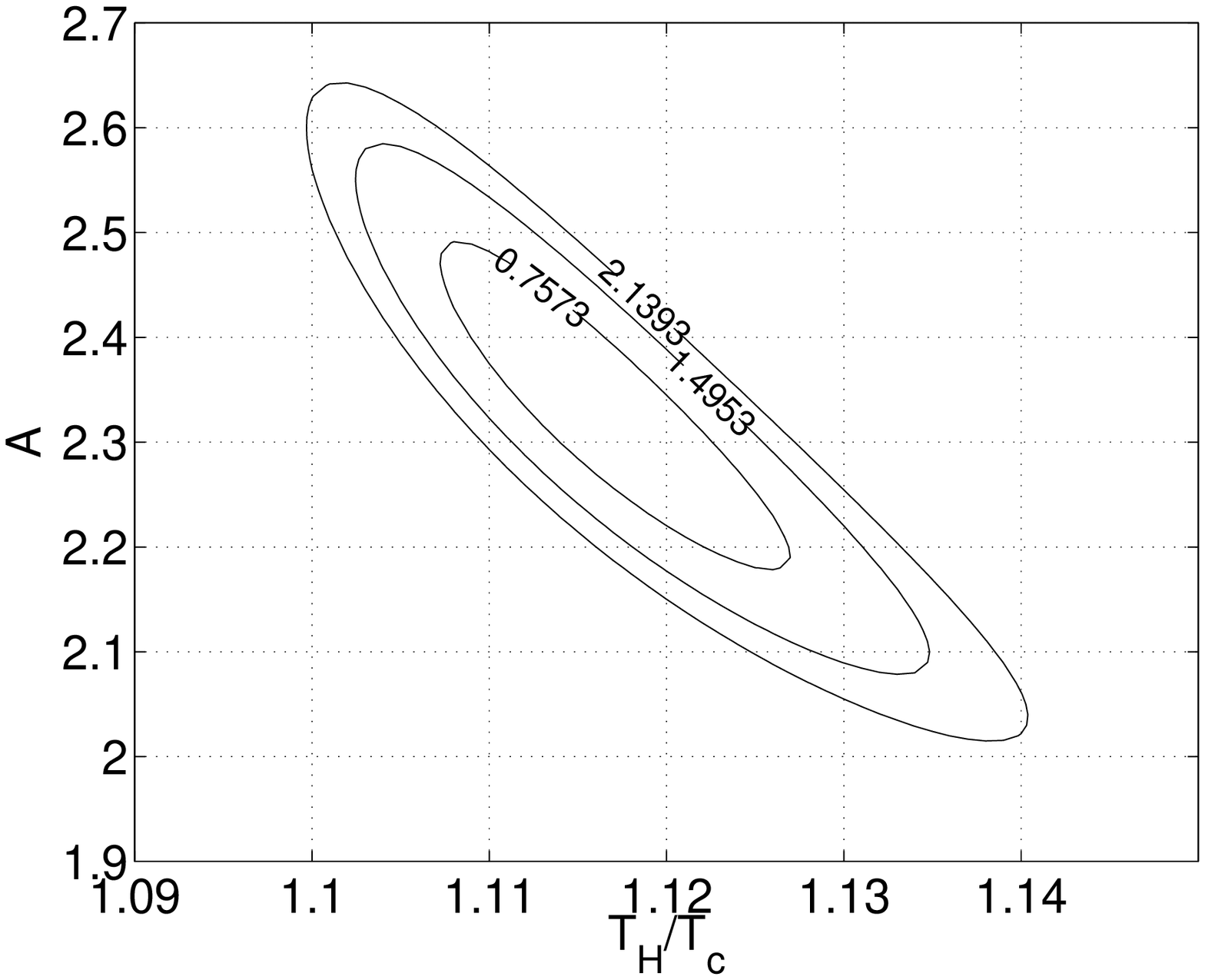}}
\vspace{0.5cm}
\caption{Confidence levels for the fits of $SU(12)$ obtained with method B, with \MF and 3D XY exponents. The 
plot on the right is for the 3D XY exponent, and on the left with a \MF exponent.} \label{fig:confidence12}
\end{figure}

\begin{figure}[htb]
\resizebox{7.5cm}{!}{\includegraphics{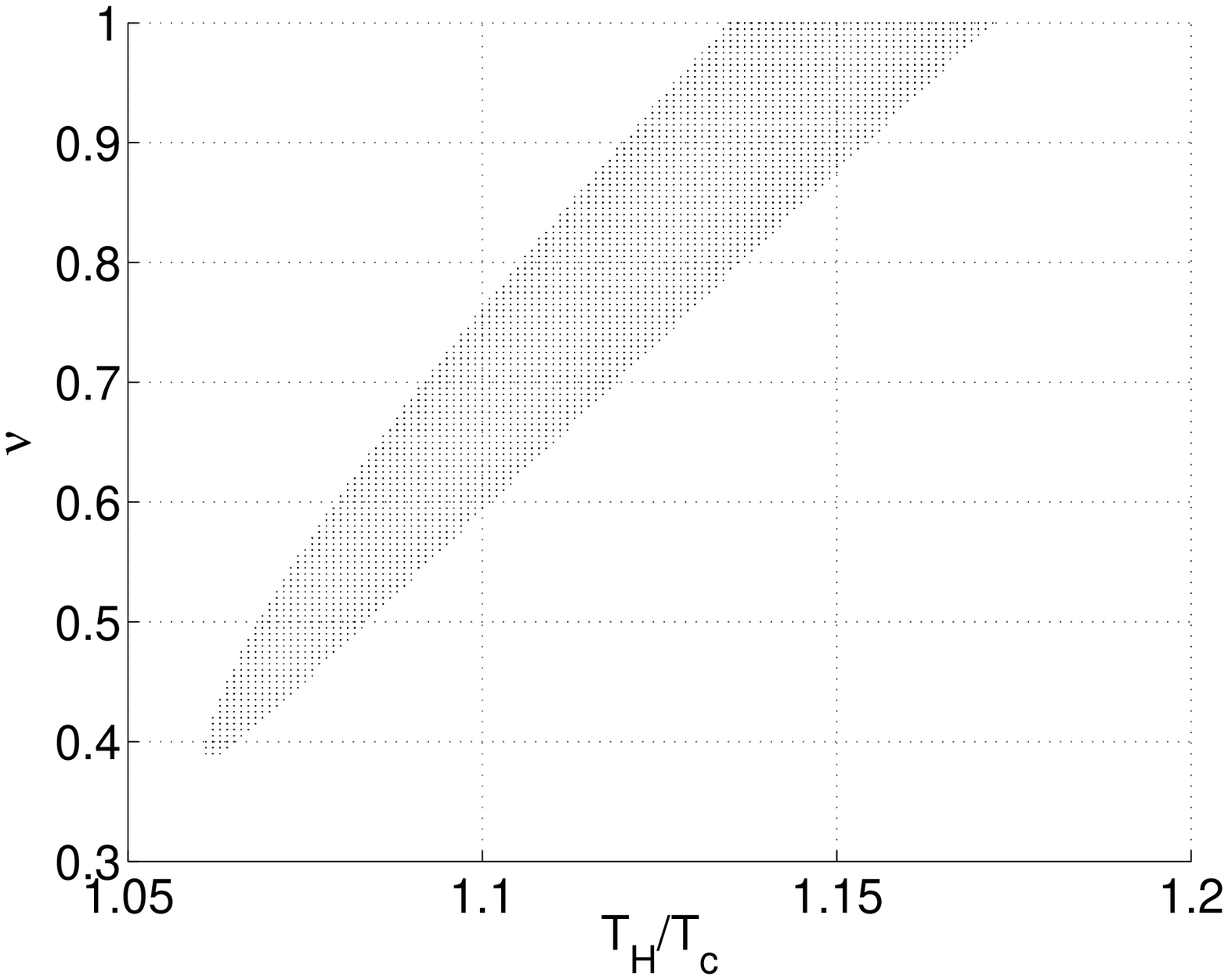}} \qquad
\resizebox{7.5cm}{!}{\includegraphics{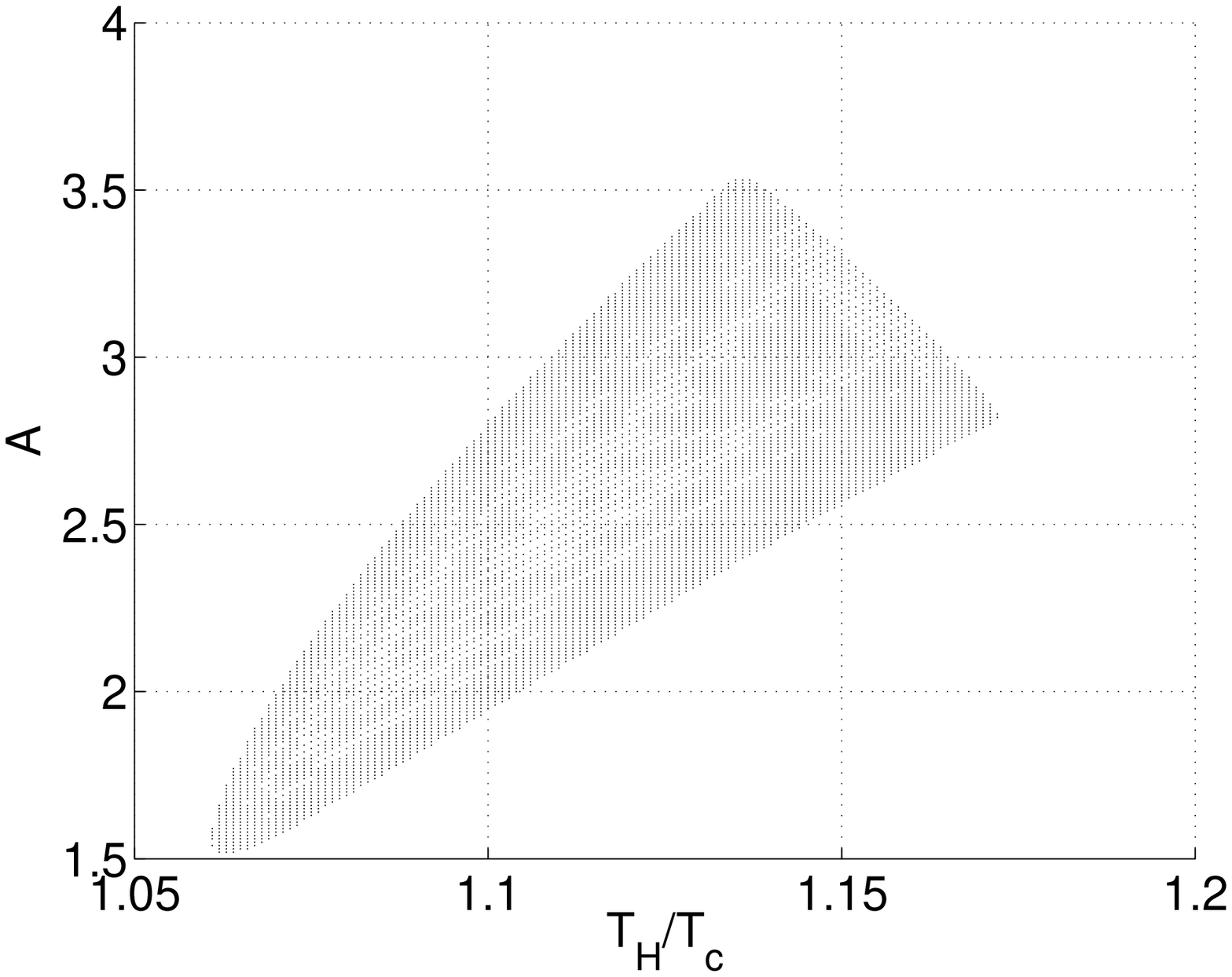}}
\vspace{0.5cm}
\caption{Two dimensional projections of a isosurface in the three-dimensional space of $(\nu,T_H/T_c,A)$ 
which represents a confidence level of $68.27\%$ for the fits of $SU(12)$ obtained with method B, with a free 
exponent.} \label{fig:confidence12free}
\end{figure}

In Figs.~\ref{fig:confidence10A}--\ref{fig:confidence12free} we present confidence levels in the fit 
parameters, for the cases where the fit is good. In the case of two parameter fits, we present contours in 
the $(A,T_H/T_c)$ plane of the $\chi^2$ per degree of freedom levels, which correspond to confidence levels 
of $68.27\%,90\%$, and $99\%$. These confidence levels are then reflected in the error estimates we give in 
the text.

When we treat $\nu$ as a free fit parameter as well, we present two dimensional projections (e.g. in 
$\nu-T_H/T_c$ space) of a volume in the parameter space of $(T_H/T_c,A,\nu)$ that corresponds to the a 
confidence level of $68.27\%$ and lower. In this case we do not quote in the text any error estimates 
together with the central values.


\section{Summary and discussion}
\label{summary}
Gathering all the statistically reliable results of $SU(10)$, we find that $T_H/T_c=1.104(6)-1.114(15)$ when 
fitting with a 3D XY exponent. When $\nu$ was treated as a free parameter fit we find central values of 
$\nu=1$ and $T_H/T_c=1.160-1.172$.
Finally when a mean-field exponent is used, one find from data obtained with method B that 
$T_H/T_c=1.087(11)$. We also find that the extrapolation
with the mean-field exponent has a higher $\chi^2$, both in the case of method A, and in the case of method B 
on a $14^35$ lattice. In particular, when we analyze both data sets together, we find that the best fit has a 
$\chi^2/{\rm dof}=1.93,1.36$ (with ${\text dof}=29$) when fitting with a mean-field, and a 3D XY exponent 
respectively. It is also interesting that the most sizable contribution to the $\chi^2$ in this case comes 
from the data point at $\beta=68.95$ (also line no.~10 in Table~\ref{table2}). Since no reasonable fit will 
go through this point (see Fig.~\ref{fig:N10}), it is quite conceivable that at this value of $\beta$ we have 
a strong statistical fluctuation. Ignoring this point gives essentially the same fitting parameters, but 
makes $\chi^2/{\text dof}=1.54,1.04$ for the mean-field and 3D XY exponents respectively. This suggests that 
the 3D XY exponent is preferable (although we still cannot exclude the mean-field exponent possibility 
completely). This preference is also be seen in Figs.~\ref{fig:confidence10AL12free}, 
and~\ref{fig:confidence10BL14free}, where we give the projections in the $\nu-T_H/T_c$ plane, of a volume in 
the $(\nu,A,T_H/T_c)$ space, that corresponds to a confidence level of $\le 68.27\%$. There we see that the 
point $\nu=0.5$ is either outside, or at the edge of this volume.

For $N=12$ the fits are better, and we find that $T_H/T_c=1.092(6)-1.116(9)$ for mean-field and 3D XY 
exponents. Both fits have a low $\chi^2$, and we again cannot rule out either. Here again the mean-field fit 
has a higher $\chi^2$ than that of the 3D XY fit. In this case, however, in view of the good $\chi^2$ values, 
the preference towards a 3D XY exponent is weaker than for $SU(10)$. Nonetheless it is interesting that when 
we perform a fit with the exponent $\nu$ as a free parameter we find $T_H/T_c=1.119$, and $\nu=0.69$ which is 
closer to the 3D XY exponent $\nu=0.6715(3)$ than to the mean-field value of $\nu=0.5$. Looking at 
Fig.~\ref{fig:confidence12free} we find that the preference of our data for $\nu=0.6715$ is indeed weaker 
here, as both exponents have a similar position in the parameter space, with respect to the shaded area.

\begin{figure}[htb]
\includegraphics[width=15cm]{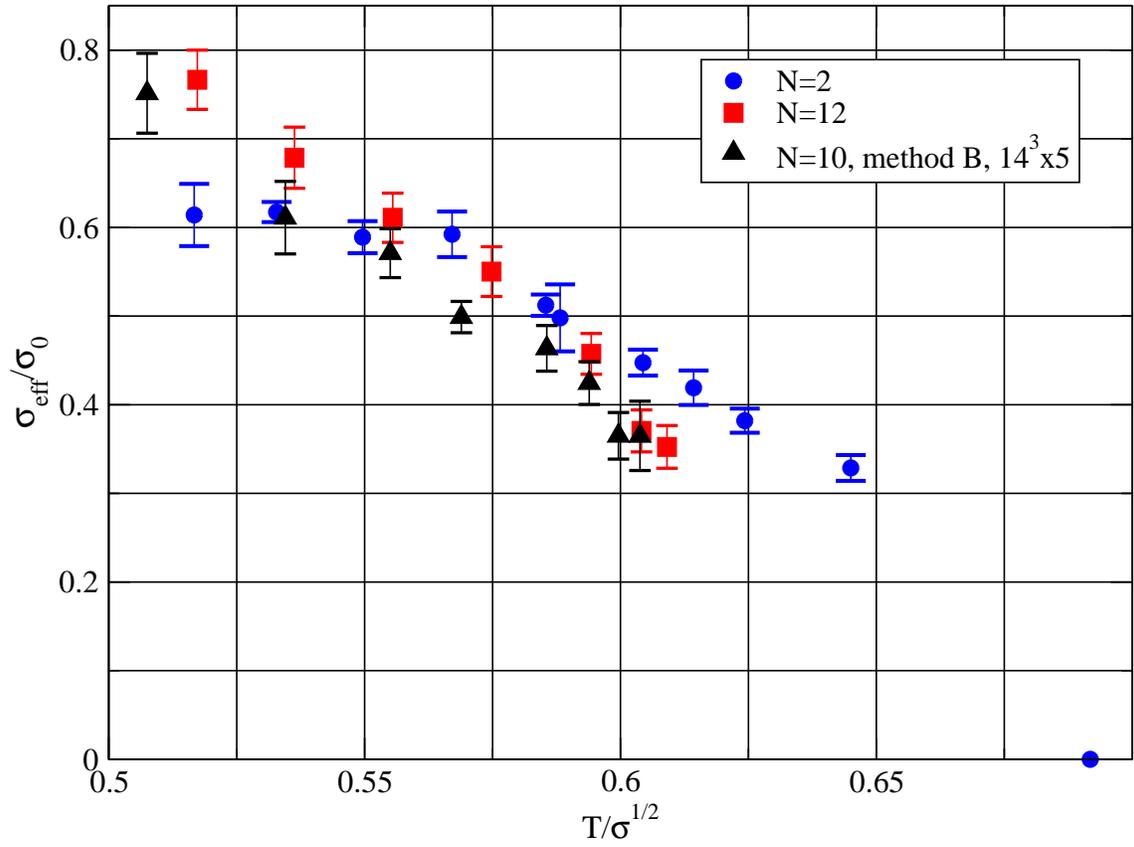}
\vspace{0.5cm}
\caption{Effective string tension for $N=2,10,12$ in units of the zero temperature string tension.}
\label{fig5}
\end{figure}

The limited statistics prevents us from making statements about the
behavior of $T_H$ as a function of $N$. This is unfortunate, since it
is of interest to know how far is $T_H/T_c$ from $1$ at
$N=\infty$. Nevertheless we obtain fitted values of
$T_H/\sqrt{\sigma}\simeq 0.62-0.68$, which is lower than
$T_c/\sqrt{\sigma}\simeq 0.7$ of $SU(2)$ \cite{McLerran:1981pb} where
the phase transition is second order, and therefore may be Hagedorn,
$T_c=T_H$, or provides a lower bound on $T_H$. To emphasize this
point we plot the $\sigma_{\rm eff}/\sigma$ for $N=10,12$
that we obtained with method B here.
For guidance we also calculate
the masses from Polyakov loops for $SU(2)$ on a $L^35$ lattice with
$L=16,20$, and $24$. The results are presented in Table~\ref{table7},
and also show the expected increase in finite volume corrections as the
temperature approaches $T_c$.
For each value of $T/\sqrt{\sigma}$ we choose to present in
Fig.~\ref{fig5} the corresponding value of $\sigma_{\rm eff}/\sigma$
calculated on the largest volume there, together with an extra point for $SU(2)$ at $T=T_c$ with $\sigma_{\rm 
eff}=0$. For the physical scale we again
use the interpolation fit in \cite{Lucini:2005vg}.
We find that for $N=10,12$ the results seem to be close, but not to
follow the $SU(2)$ results.

\begin{table}[htb]
\caption{Masses from Polyakov loops on $L^35$ lattices for $N=2$.
\label{table7}}
\begin{ruledtabular}
\begin{tabular}{ccccc} \hline
\multicolumn{5}{c}{loop mass: SU(2)} \\ \hline
$\beta$ & $16^3 5$ & $20^3 5$ & $24^3 5$ & $a\sqrt{\sigma}$ \\ \hline
2.28   & 0.472(11) & 0.476(13) &  0.460(26) &  0.3871 \\
2.29   & 0.439(9)  & 0.435(8)  &   - & 0.3754\\
2.295   & 0.417(22)  & -  &    0.383(22)&  0.3639\\
2.30   & 0.373(11) & 0.390(12) &  - & 0.3527\\
2.31   & 0.359(8)  & 0.335(8)  &   0.368(16) & 0.3417\\
2.32   & 0.303(8)  & 0.299(9)  & 0.299(7)  &  0.3400\\
2.3215   & 0.290(19)  & -  &    0.288(22) & 0.3309\\
2.33   & 0.274(8)  & 0.252(8)  & 0.245(8) &  0.3256\\
2.335   & 0.252(11)  & -  &    0.222(10) & 0.3204\\
2.34   & 0.217(8)  & 0.205(7)  & 0.196(7) &  0.3101\\
2.35  & 0.161(6)  & 0.158(7)  & 0.158(7)  & 0.2891\\ \hline
\end{tabular}
\end{ruledtabular}
\end{table}

Finally we find interesting the fact that the Nambu-Goto action, gives
rises to a Hagedorn behavior at similar temperatures given by
\begin{equation}
T^{\text{NG}}_c/\sqrt{\sigma}\simeq 0.691/\sqrt{c}, \label{Nambu-Goto}
\end{equation}
where $c$ is the central charge \cite{Meyer:2004hv} and equals unity
in the usual Nambu-Goto model. Applying this
formula to our values of $T_H$ we give the central
charge values listed in Table~\ref{table6}.

A determination of the proper universality class is an important
issue, and may teach us how the scaling region behaves with $N$
(if at $T_H$ the system behaves like in a second order transition).
As mentioned in the introduction, this question was studied numerically in
\cite{Chandrasekharan:2004uw} for the strongly-coupled gauge theory
(with quarks included), and analytically in \cite{me_strong} where the
scaling region for chiral restoration was seen not to
change with $N$. In our context the similar question can
be approached for deconfinement in the continuum of the pure gauge
theory. Unfortunately, despite
the mild preference towards the 3D XY model, discussed above, we
currently cannot rule out unambiguously any of the universality
classes.
To distinguish which one is actually correct one must
approach $T_H$ closer, and increase the statistics. In fact, as
discussed in the introduction, if the Hagedorn transition is second
order, we expect that the
critical region shrinks with increasing $N$, and that only when $(T_H-T)\stackrel{<}{_\sim}
1/N^4$, one will see the nontrivial critical behavior of a 3DXY
model. This is however hard because tunneling
configurations become more probable, and finite volume effects
(although relatively small at larger
values of $N$) become more important.
Nonetheless when we perform a fit with the critical exponent $\nu\in
[0,1]$ as a free parameter we find that
for $SU(12)$, the best fit result for $\nu$ is closer to the 3D XY exponent than to the mean-field value.

We believe that a more thorough investigation (with larger statistics)
would render the understanding of the proper universality class, and
indeed all the other issues discussed above, much clearer, and the
large--$N$ limit of the gauge theory better understood. However,
considering its current numerical cost we postpone it to future
studies. A different route to approach this question is to study
variants of the pure gauge theory, such as adding scalar fields. Depending on the couplings added, the phase 
transition can
become second order \cite{Aharony:2005bq}, and one can study the
adequacy of large-$N$ mean-field techniques close to second order
phase transition in a thermodynamically stable
phase.

\begin{acknowledgments}
Our lattice calculations were carried out on PPARC
and EPSRC funded computers in Oxford Theoretical
Physics. BB was supported by a PPARC. We thank Ofer Aharony, John Cardy, Simon Hands, Maria Paola Lombardo, 
and Benjamin Svetitsky for useful discussions and remarks.
\end{acknowledgments}

\bibliography{Hagedorn}

\end{document}